\documentstyle[12pt]{article}

\global\arraycolsep=1pt
\newcommand{\beq}{\begin{equation}}
\newcommand{\eeq}{\end{equation}}
\newcommand{\beqa}{\begin{eqnarray}}
\newcommand{\eeqa}{\end{eqnarray}}
\newcommand{\CR}{\nonumber \\}

\newcommand{\del}{\partial}

\newcommand{\trace}{\hbox {Tr}~}

\def\LG{{  \cal G}}
\def\LA{{   \cal A}}
\def\LF{{ \cal F}}
\def\LM{{ \cal M}}
\def\LO{{\cal  O}}
\def\L{\Lambda}

\def\dsla{\not\!\partial}
\def\tensor{\otimes}
\def\tod{\stackrel{d}{\longrightarrow}}
\def\toDA{\stackrel{{D_A}}{\longrightarrow}}
\def\O{\Omega}
\def\LH{{\cal\bf H}}

\def\N{{1\over {8\pi^2}}}

\def\z{\bar z}
\def\m{\mu}
\def\n{\nu}
\def\pa{\partial}
\def\LO{{\cal O}}
\def\o{\omega}

\def\D{{\it D}}

\begin{document}

\renewcommand{\thefootnote}{\fnsymbol{footnote}}

\begin{titlepage}
\begin{flushright}
         LPTHE-9707\\
   hep-th/9704167
\end{flushright}
\begin{center}
{\Large \bf Special Quantum Field Theories \\ In  Eight And Other
Dimensions   }
\lineskip .75em
\vskip 3em
\normalsize
{\large  Laurent Baulieu}\footnote{email address:
baulieu@lpthe.jussieu.fr}   \\
{\it LPTHE, Universit\'es Paris VI - Paris VII, Paris,
France}\footnote{
URA 280 CNRS,
4 place Jussieu, F-75252 Paris Cedex 05, France.} \\
\vskip 1 em
{\large Hiroaki Kanno}\footnote{email address:
  kanno@math.sci.hiroshima-u.ac.jp} \\
{\it Department of Mathematics, Faculty of Science,
Hiroshima University, \\
Higashi-Hiroshima 739, Japan}\\
\vskip 1 em
{\large I. M. Singer}\footnote{email address:
ims@math.mit.edu} \\
{\it  Department of Mathematics, MIT, Cambridge, USA}
\end{center}
\vskip 1 em
\begin{abstract} We   build  nearly topological quantum field theories in
various dimensions. We    give  special attention to the case of 8 dimensions
for which we first consider theories depending only on Yang-Mills fields. Two
classes of gauge functions exist which correspond   to the choices of two
different holonomy groups in $SO(8)$, namely $SU(4)$ and $Spin(7)$. The choice
of $SU(4)$ gives a quantum field theory for a Calabi-Yau fourfold.  The
expectation values for the observables  are formally holomorphic Donaldson
invariants. The choice of $Spin(7)$ defines  another eight dimensional theory
for a Joyce manifold which could be of relevance in $M$- and $F$-theories.
Relations to the eight dimensional supersymmetric Yang-Mills theory are
presented. Then, by   dimensional reduction, we obtain other theories, in
particular a four dimensional one  whose   gauge conditions are  identical to
the non-abelian Seiberg-Witten equations. The latter are thus related to  pure
Yang-Mills self-duality equations in 8 dimensions as well as to the  N=1, D=10
super Yang-Mills theory. We also exhibit a theory that couples 3-form gauge
fields to the second Chern class in eight dimensions, and interesting theories in other dimensions.
\end{abstract}

\end{titlepage}
\renewcommand{\thefootnote}{\arabic{footnote}}
\setcounter{footnote}{0}


\section{Introduction}

\renewcommand{\theequation}{1.\arabic{equation}}\setcounter
{equation}{0}

Topological quantum field theory (TQFT), or more
specifically, cohomological
quantum field theory has been extensively studied in two,
three and four
dimensions. (See e.g. \cite{BBRT}, \cite{CMR} and references
therein.) In this
article we show that theories which are almost topological
also exist in
dimensions higher than four. We call them BRSTQFT's instead
of TQFT's.
We give special attention to the case of Yang-Mills fields
in eight dimensions.

A BRSTQFT relies on a Lagrangian which contains as many
bosons as fermions,
interconnected by a BRST symmetry. The Lagrangian density is
locally a sum of
d-closed and BRST-exact terms. Starting from classical
 \lq\lq topological\rq\rq\ invariants,
the most crucial point
 in the construction of the BRSTQFT is the determination of
gauge fixing
conditions, enforced in a BRST invariant way. In the weak
coupling expansion,
one interprets the theory as exploring through path
integrations all quantum
fluctuations around the solutions to the gauge conditions.
This provides,
eventually, an intuitive way to study the moduli problem
associated
with the choice of gauge fixing conditions,
by computing Green functions defined from the BRST
cohomology. Generally, one must distinguish between  the
ordinary gauge fixing
conditions for the ordinary gauge degrees of freedom of
forms and the gauge
covariant  ones which occur when one gauge-fixes a
"topological" invariant
(i.e, constant on a Pontryagin sector of gauge fields).  A
 BRSTQFT  can often be untwisted 
into  a  Poincar\'e supersymmetric theory; we give more examples 
in this paper.
BRSTQFT's are microscopic
theories, in the sense that in principle they provide the
fundamental fields to
study (almost) topological properties. We ask: are their
infrared limits
describable by effective theories, following the ideas of
Seiberg and Witten? 

In four dimensions Donaldson \cite{DO} used the moduli space of
anti-self-dual fields to describe invariants of four manifolds.
Witten \cite{Wi1} interpreted these invariants as
observables in a topological quantum field theory,
twisted $N\!=\!2$ supersymmetric Yang-Mills.
Baulieu and Singer \cite{BS} noted that this TQFT could be
obtained
from a topological action by the BRST formalism with
covariant
gauge functions which probe the moduli space of 
anti-self-dual fields.  
In this paper, we apply this formalism to higher dimensional
cases of self duality; M-theory, F-theory, 
and low energy limits of string theory have increased the interest in QFTs 
in dimension greater than 4.

Over a decade ago, Corrigan et al \cite{CDFN} classified
the cases in which the
self-duality equation for Yang-Mills fields  in 4 dimensions could be
generalized
to higher dimensions. See also  Ward \cite{WARD}.
Solutions to these equations are higher dimensional
instantons
\cite{FN1} \cite{FN2}.  The generalizations in eight dimensions depend on 
having the holonomy group reduced from $SO(8)$ to ${{\em Spin}(7)}$  or
$SU(4).$  
See Salamon \cite{SALA} for background on special holonomy groups.

The third author (IMS) learned about self-duality in eight dimensions 
for Einstein manifolds and fields associated to the spin bundle 
from Eric Weinstein in 1990.  Weinstein constructed special instantons, 
computed the dimensions of the corresponding moduli space, 
and noted the importance of ${\em Spin} (7)$ and $SU(4)$. For this, and
more, see \cite{Weinstein}.

The geometry for manifolds with holonomy ${\em Spin}(7)$ can be found 
in Joyce \cite{Jo1}.  For holonomy $SU(4),$ the holomorphic extension of
 Donaldson Theory is being developed by Donaldson, Joyce, Lewis, and Thomas
 at Oxford.  Their program for extending results in two, three and four
dimensions 
from the real to the complex case is sketched in Donaldson and Thomas
\cite{DOT}.

 In the first
part of this paper we describe two eight dimensional Yang-Mills quantum field
theories that reflect the eight dimensional self duality
equations
found in \cite{CDFN}; we use the geometry developed by the above-mentioned 
authors to construct the quantum field theory.
These theories cannot be called  topological for they depend on some
geometrical
structure of the manifold $M_8.$  For want of a better term,
we have called them BRST quantum
field theories (BRSTQFT), because they are constructed by
starting with a
topological action and using the BRST formalism
with covariant gauge functions that
again probe the moduli space of these new anti-self-dual
fields.

When the holonomy group is ${\em Spin}(7) \subset SO(8),$ we call $(M_8,g)$
a Joyce manifold.  Section 2.1.1 gives the geometry needed to 
construct the BRSTQFT of 2.1.2, which is in turn described geometrically in
2.1.3.
Section 2.2 gives a parallel discussion of the holomorphic case, 
i.e., when the holonomy groups is $SU(4).$  We compare the two cases 
in  section 
2.3.
We point out in Section 2.4 that the J-case is a twist of
$D\!=\!10$, $N\!=\!1$ supersymmetric Yang-Mills theory (SSYM)
dimensionally reduced  to
$D\!=\!8$. Since supersymmetries for a curved manifold 
require covariant constant
spinors, there is one remaining supersymmetry; we explain 
 its relation to the
topological BRST symmetry.

Having defined pure Yang-Mills BRSTQFT in eight dimensions,
we introduce
a different theory in section 3 which couples an uncharged 3-form
gauge field $B_3$ to the Yang-Mills field $A$.
We propose as covariant gauge conditions of the coupled
systems, the pair of equations
\beqa
F_A &=& * \Omega\wedge F_A~,  \CR
\trace ( F_A \wedge F_A) &+& dB_3+*dB_3= 0~, \label{instap}
\eeqa
where $\Omega$ is a background closed 4-form. One must be careful here; 
$B_3$ is not an ordinary 3-form
and $dB_3$ is
not its differential. Rather, $B_3$ is locally defined, up
to an exact 3-form
so that $dB_3$ stands for a closed 4-form. (See the
discussion in section 3).
\cite{GN} and \cite{ivano} give  an explicit solution of eq.~(\ref{instap}) for
$M_7 \times
{\bf R}.$

Section 4 discusses other dimensions.   When $M_{12}$ is a Calabi-Yau 6-fold,
one can define BRSTQFT's  and we do so.
We reduce our 8D theories to 6D and 4D in sections 4.2 and
4.3, respectively.
The H case reduction can be obtained directly on a Calabi-Yau 3-fold by
a modification of the methods in section 2.2.

The reduction to 4D is particularly interesting. On the one
hand we get a twisted $N\!=\!4$ SSYM of Vafa and Witten \cite{VW}. In fact,
the H, J cases and the case of $M_7$, holonomy $G_2$ theory,
reduced to 4D, give the three twists of $N\!=\!4$ SSYM. 
On the other hand we also get the nonabelian Seiberg-Witten theory.
Thus    there is a relationship between $N\!=\!4$ SSYM
and nonabelian SW theories. The latter theory is obtained from  the 
eight dimensional J theory, with its octonionic structure; the former
is obtained from
 the $N\!=\!1, D\!=\!10$ SSYM theory, by ordinary dimensional
reduction. 
The direct link between the  $D\!=\!10$ SSYM theory and the J theory
is that 
the $N\!=\!1,  D\!=\!10$ SSYM theory gives by dimensional reduction the  
$N\!=\!1,  D\!=\!8$ SSYM which can be identified 
with the J theory by a simplest twist, specific to 8
dimensions,  which interchanges 
vectors and spinors (Section 2.4).


\section{Pure Yang-Mills 8 dimensional case}

\renewcommand{\theequation}{2.\arabic{equation}}\setcounter
{equation}{0}

 The 4 dimensional Yang-Mills TQFT can be obtained by the BRST
formalism.
Starting with $p_1= \N \trace F\wedge F$, one gauge-fixes
its invariances with three covariant gauge conditions
and one Feynman-Landau gauge condition that probe the moduli
space of
self-dual curvature fields \cite{BS}.
These gauge conditions are enforced in a BRST
invariant way, by using the 4 gauge freedom of local general
infinitesimal
variations of the connection $A_\mu$. Put mathematically, we
get an elliptic
complex $0\to\L^0\tod \L^1 \tod
\L^2_+\to 0$, tensored with a Lie algebra $\LG$.

In this section we extend this scheme to 8 dimensions when the
holonomy group
in $SO(8)$ is either $SU(4)$ (the case of a Calabi-Yau 4-fold)
or $Spin (7)$ (the case of a Joyce
manifold). The 4-D self duality equations must be
generalized to
\beq
\lambda F^{\mu\nu}=  \frac{1}{2}T^{\mu\nu\rho\sigma} F_{\rho\sigma}~,
\label{basic}\eeq
where $\lambda$ is a constant (an eigenvalue) and 
$T^{\mu\nu\rho\sigma}$ is a totally antisymmetric tensor 
which is generally not invariant under 
general $SO(D)$ transformations. Rather it is invariant
under a subgroup of $SO(D)$. Corrigan et al \cite{CDFN} classified the 
possible choices of  $T^{\mu\nu\rho\sigma}$ up to eight
dimensions,
where two solutions $T$ are singled out.
Indeed, for these cases, the space of 2-forms
$\L^2$ decomposes into a direct sum and one can thus replace
the self-duality
condition in 4 dimensions by the condition that the
curvature fields lie in an
appropriate summand. The elliptic complex above has an 8-D
counterpart:
$0\to\L^0\tod \L^1 \tod
P_+(\L^2)\to 0$.
 Moreover, in each case, there is a closed 4-form $\Omega$
and one can replace
$p_1$ by
 \beq
\N \Omega \wedge \trace (F\wedge F)~.
\eeq
Since $\int_X \Omega \wedge \trace (F\wedge F)$ is
independent of the gauge field $A$ and since the new
elliptic complex
implies that the number of gauge covariant gauge functions
plus Feynman-Landau type gauge
condition is eight, one can use the BRST formalism
to introduce new (ghosts and ghosts of ghosts) fields and an
invariant action.
The theory is not topological, because it depends on the
reduction
of the holonomy group.
In the case of the $SU(4)$ reduction, one predicts that the
expectation
value of the observables depends on the holomorphic
structure of $X$,
but not on the choice of the Calabi-Yau metrics.
We call these theories BRSTQFT's. We will say
the BRSTQFT is of type J for $Spin (7)$ and of type H
for $SU(4)$.
We will analyze each case.
They differ in a subtle way from the point of view of BRST
quantization. In the type H case one has 6 independent
real covariant gauge
conditions which can be seen as 3 complex 4-D self-duality
conditions. We can
complete them by a {\it complex} Landau type gauge condition
which counts for the 2 missing gauge conditions allowed by
the eight
freedom in deforming the Yang-Mills
field. In the type J case one has 7 independent real
equations which we can
complete by the usual (real) Landau gauge condition. In the
former case one has
thus a complexification of all ingredients of the 4-D case.
In the latter
case all
fields are real, and the situation is quite like the 4-D
case, with the change of the quaternionic structure of the self duality
equations 
in four dimensions into an octonionic one in eight dimensions.

The action we consider will be the BRST invariant gauge
fixing of the
topological invariant
\beq
S_0~=~\frac{1}{2} \int_{M_8}~ \Omega \wedge \trace ( F
\wedge F)~,
\eeq
where $\Omega$ is a fixed closed four form adapted to each case.
Depending on the case,
we will have six or seven  covariant gauge fixing conditions of
the type of eq.~(2.1),
that we will denote as $\Phi_i =0$, $1\leq i \leq 6 \
\rm{or}\ 7$.
That we get an action containing a Yang-Mills part relies on
the identity
\beq 
a  \sum_{i}\trace (\Phi_i \Phi_i)\cdot~(vol)~=
~-S_0 + ~\trace (F \wedge * F)~, \label{square}
\eeq
\noindent
where  $a$ is a  positive real number (one has different
decompositions in the J and H cases).
$(vol)$ stands for the volume form.
 The last term is the  
action density for the Yang-Mills theory.
Hence a solution to $\Phi_i=0 $  gives a
stationary point of the eight dimensional Yang-Mills theory.
For this reason, the equations
$\frac{1}{2}T^{\mu\nu\rho\sigma} F_{\rho\sigma}~=~\lambda
F^{\mu\nu}$,
deserve to be called the instanton equation. Notice that
one has the correspondence
$ \Omega_{\mu\nu\rho\sigma} =\epsilon _{\mu\nu\rho\sigma
\alpha\beta\gamma\delta}T^{\alpha\beta\gamma\delta}$.
By adding to $S_0$ a BRST exact term which generates among
other terms $\sum_{i}(\Phi_i \Phi_i)$, we will
thus replace the 'topological' invariant $\Omega \wedge \trace
(F \wedge F) $
by the standard Yang-Mills Lagrangian $\trace (F \wedge *
F)$ plus ghost terms, which constitute the action of the BRSTQFT
theory. As explained  earlier, the term BRSTQFT seems to us more
 appropriate than the term TQFT for the resulting theory.
Obviously, the remaining gauge invariances must be gauge
fixed,
which will be done in the same spirit, as in \cite{BS}.

\subsection {Type J case: Joyce manifold}

\subsubsection{Geometrical setup}

 Recently it has been proposed that the $7$ dimensional and
the $8$ dimensional Joyce manifolds provide a
compactification
to four dimensions of $M$-theory and
$F$-theory, respectively \cite{Vaf} \cite{PT} \cite{Ach}.
We consider here the $8$ dimensional  case and call
a Joyce manifold an eight dimensional manifold with
$Spin(7)$ holonomy \cite{Jo1} \footnote{ There is another
class of Joyce
manifolds in seven dimensions \cite{Jo2}. Its holonomy is the
exceptional group
$G_2$. Both classes of Joyce  manifolds have been
studied in superconformal field theory \cite{SV} \cite{Fig}. }. Then
$Spin(7)$ acting on
$\L^4(M_8)$, the space of 4-forms, leaves invariant a self-dual 4-form
$\Omega \neq 0$. Further, $\Omega $ is covariantly constant
and
hence closed. The space of 2-forms $\L^2(M_8)$ splits into
$\L^2_{21} \oplus \L_+^2$ with $ {\rm \dim\ } _{\bf R} \L_+^2 =7$.
One can see this by noting that $\L^2\simeq so(8)$ and that
$\L^2_{21} \simeq$
 Lie algebra of $Spin(7) \subset so(8)$. The splitting can
also be obtained as
follows: let $T$ be the operator on $\L^2$ given by
$\tau\to *(\Omega\wedge \tau)$.
Then $T$ is self adjoint with eigenvalues $+1$ and $-3$, when
$\O$ is scaled.
Its eigenspaces are $\L^2_{21}$ and $\L^2_{+},$ respectively.
The ordinary anti-self-dual Yang-Mills
fields in four dimensions are now to be replaced by
$(P_+F_A)=0$,
where $P_+$ is the projection of $\L^2$ onto $\L^2_+.$
We next discuss the linearization of this equation.

Let $S_M^+$ and $S_M^-$ (that is, ${\bf 8_s}$ and ${\bf 8_c}$
in another
notation) denote the chiral and antichiral real (Majorana)
spinors for
$M_8$ ($M_8$ is simply connected and has a unique spin
structure).
Then the representation of $Spin (7)$
on $S_M ^+$ is the direct sum ${\bf R}\oplus {\bf V}$
(that is, ${\bf  8_s} = {\bf 1} \oplus {\bf 7}$).
Let $\zeta$ be a covariantly constant spinor
field of norm $1$ giving the splitting of $S_M^+$.
The representation of $Spin(7)$  on $S_M^-$ is
irreducible. Since $S_M\tensor S_M$ is isomorphic to forms,
tensoring by $\zeta$
 identifies spinors with forms. For example, $\L^2(S^+_M
)\simeq \L^2(M_8)$;
so $\L^2(S^+_M ) = \L^2({\bf  R}\oplus {\bf  V}) ={\bf V}
\wedge {\bf V}+ \zeta \tensor
{\bf V}$ gives the splitting into $\L^2_{21}\oplus
\L^2_{+}$.
Further $\zeta
\tensor S_M^-$ can be identified with $\L^1(M_8)$, that is,
${\bf 8_v }$.
We conclude that the sequence
$0\to\L^0\tod \L^1{\stackrel{{P_+ d}}{\longrightarrow} }
\L^2_+\to 0$ is an elliptic sequence and
 $( P_+d+d^* ):\L^1\to \L^2_+\oplus \L^0$ is the Dirac operator
$\dsla: S_M^-\to S_M^+,$ after the
identification of spinors with  forms due to $\zeta$.

If $P$ is a principal bundle over $M_8$ with a compact gauge
group $G$, 
we can couple forms to its Lie algebra $ \LG$ by a vector
potential $A$.
We have the sequence
$0\to \L^0\tensor \LG\toDA \L^1 \tensor \LG
{\stackrel{{P_+D_A}}
{\longrightarrow } }\L^2_+
 \tensor \LG\to 0$ which is elliptic when
$P_+D_A^2=0$, i.e. when $P_+F_A =0$. 
(Here  we have identified the 
Lie algebra $\LG $ with the adjoint
Lie algebra bundle over $M_8$.) 
In general, $P_+D_A+D_A^*=\not\!\! D _A \ :
\L^1\tensor \LG\to \L^2_+\tensor \LG+ \L^0\tensor \LG $ is
elliptic.
The index of the operator is the virtual dimension of
the moduli space $\LM_J$ of solutions to the nonlinear equation $
P_+ F_A=0$, modulo gauge transformations.

To make contact with the next section, let us remark that
$P_+  F_A=0$
determines, in the case of a pure Yang-Mills BRSTQFT,
the relevant gauge covariant gauge
conditions shown in eq.~(2.1), while $D^*_A$ is
the operator related to the Landau-Feynman
 gauge condition of ordinary gauge degrees of freedom.
\def\o{\omega}

More precisely, the BRSTQFT that will be determined shortly
is
the gauge fixing by BRST techniques of
$S_0[A]=\int_{M_8} \O \wedge \trace (F\wedge F)$. The latter
is independent
of $A$, since it is $8 \pi^2 \O \cup p_1(P)$ which only depends
on the topological charge of $A$.

The way one gets the Yang-Mills action from the gauge fixing
of
an invariant is the consequence of the following.
If $\o$ is an element of $\L^2$, let $\o_-$ and
$\o_+$ be its components on $\L^2_{21}$ and $\L^2_+$. Then
$\parallel\!\o\!\parallel^2=\parallel\!\o_+\!\parallel^2
+\parallel\!\o_-\!\parallel^2$,  
$\langle \o_+, \o_- \rangle =0,$ while
\beqa\label{ims1}
\O \wedge F\wedge F &=& \O \wedge (F_++F_-)\wedge (F_++F_-)\CR
 &=& \O\wedge F_+ \wedge F_+ +\O \wedge F_- \wedge F_-+\O
\wedge F_- \wedge F_++\O
\wedge F_+ \wedge F_-\CR 
&=& -3* F_+ \wedge F_+ +* F_-
\wedge F_- +* F_- \wedge F_+
  -3* F_+ \wedge F_-~.
\eeqa
Thus
\beq
\int_{M_8} \trace \left (\O \wedge F\wedge F\right )=
\parallel\! F_-\!\parallel^2-3\parallel\! F_+\!\parallel^2~, \label{pnorm}
\eeq
and
\beq\label{ims3}
 \parallel\! F _A\!\parallel^2 =\int_{M_8} \trace \left (\O \wedge F_A\wedge
F_A\right )+4 \parallel\!
F_+\!\parallel^2 ~.  \label{norm}
\eeq
$\Omega \wedge \Omega$ orients $M_8$ and is the volume element.  
Given the topologial sector, we choose $\Omega$ so that $\int_{M_8} \trace
(\O \wedge F_A\wedge F_A)\geq 0$. Then
 $F_+=0$, minimizes the action $ \parallel\! F _A\!\parallel^2.$

To write the BRSTQFT action in the physicist notation,
we have to be more explicit.
In terms of an orthonormal basis, the self-dual four form is 
\beqa
\Omega &=& e_1 \wedge e_2 \wedge e_5 \wedge e_6 + e_1
\wedge e_2 \wedge e_7 \wedge
e_8+ e_3 \wedge e_4 \wedge e_5 \wedge e_6 \CR & & + e_3
\wedge e_4 \wedge e_7
\wedge e_8 + e_1 \wedge e_3 \wedge e_5 \wedge e_7 - e_1
\wedge e_3 \wedge e_6
\wedge e_8 \CR & & - e_2 \wedge e_4 \wedge e_5 \wedge e_7 +
e_2
\wedge e_4 \wedge
e_6 \wedge e_8 - e_1 \wedge e_4 \wedge e_5 \wedge e_8 \\ & &
- e_1
\wedge e_4
\wedge e_6 \wedge e_7 - e_2 \wedge e_3 \wedge e_5 \wedge e_8
- e_2
\wedge e_3
\wedge e_6 \wedge e_7 \CR & & + e_1 \wedge e_2 \wedge e_3
\wedge e_4 + e_5 \wedge
e_6 \wedge e_7 \wedge e_8~, \nonumber
\eeqa
where $e_i~(i=1, \ldots, 8)$ are vielbein fields.

The operator $T$ defined above can be written
as the following $Spin(7)$ invariant fourth rank
antisymmetric tensor
\beq
T^{\mu\nu\rho\sigma}~=~\zeta^T \gamma^{\mu\nu\rho\sigma}
\zeta~,
\eeq
where
 $\gamma^{\mu\nu\rho\sigma}$ is the totally
antisymmetric product
of $\gamma$ matrices for the $SO(8)$ spinor representation;
\beq
\gamma^{\mu\nu\rho\sigma}~=~\frac{1}{4!}
\gamma^{[\mu}\gamma^{\nu}\gamma^{\rho}\gamma^{\sigma]}~,
\eeq
and $\zeta$ is the covariantly constant spinor introduced above 
to identify spinors with forms.
This gives another component representation of the four form
$\Omega$.
To repeat the first paragraph
of this section in terms of the   fourth rank tensor
$T^{\mu\nu\rho\sigma}$, we define an analogue of the
instanton equation
on the Joyce manifold \cite{CDFN};
\beq
F^{\mu\nu}~=~\frac{1}{2} T^{\mu\nu\rho\sigma}
F_{\rho\sigma}~,
\quad{\rm i.e.} \quad F \in \L^2_+\label{Jinst}
\eeq
The curvature 2-form $F_{\mu\nu}$ in 8
dimensions has 28 components, whose $Spin(7)$ decomposition
is
${\bf 28 = 7 \oplus 21}$. (
This is made explicit by the eigenspace decomposition of the
action of
$\frac{1}{2} T^{\mu\nu\rho\sigma}$  in eq.~(\ref{basic})
with the eigenvalues
$\lambda = -3$ and $\lambda = 1$.)

Eq.~(\ref{Jinst}) can be written as seven independent
equations,
showing that the curvature has no components in the former
subspace
which is 7-dimensional
\beq
F_{8i}=c_{ijk} F_{jk}, \quad 1 \leq i,j,k \leq 7~.
\label{octo}
\eeq
 Eq.~(\ref{octo}) makes the octonionic structure explicit.
Indeed, the $c_{ijk}$
are the structure constants
 for octonions\footnote{If we decompose the octonions into its one
dimensional real part and 7 dimensional
 imaginary part, $R^7,$ then $*_7(\Omega|_{R^7})$ is a 3-form $\alpha$
which determines Cayley multiplication on $R^7$ by $\alpha  (z,y,z) 
=<x,y,z>.$} \cite{GG} and the eight dimensional tensors
$T_{\mu\nu\rho\sigma}$ can be written as \footnote{ In the four
 dimensional case   one has similar equations,
with the indices $i,j,k$ running from 1 to 3. Then the
coefficients
$c_{ijk}$ are
the structure constants for quaternions. The holomorphic H
case that we will
shortly analyze is thus a theory with a complexified
quaternionic structure.}
\beqa
T_{8ijk}&=&c_{ijk}, \quad 1 \leq i,j,k \leq 7\CR T_{lijk}&=
&{1\over 24}
\epsilon^{lijkabc}c_{abc}, \quad 1 \leq i,j,k,l \leq 7~.
\eeqa
Notice that by construction, the $T_{\mu\nu\rho\sigma}$ are
self-dual objects in 8 dimensions.  Computed explicitly, eq.
(\ref{Jinst}) is
\beqa
\Phi_1 &\equiv& F_{12} + F_{34} + F_{56} + F_{78} = 0~, \CR
\Phi_2 &\equiv& F_{13} + F_{42} + F_{57} + F_{86} = 0~, \CR
\Phi_3 &\equiv& F_{14} + F_{23} + F_{76} + F_{85} = 0~, \CR
\Phi_4 &\equiv& F_{15} + F_{62} + F_{73} + F_{48} = 0~,
\label{gaugefix} \\
\Phi_5 &\equiv& F_{16} + F_{25} + F_{38} + F_{47} = 0~, \CR
\Phi_6 &\equiv& F_{17} + F_{82} + F_{35} + F_{64} = 0~, \CR
\Phi_7 &\equiv& F_{18} + F_{27} + F_{63} + F_{54} = 0~.
\nonumber
\eeqa
In this form, the gauge functions are ready to be used to
define the BRSTQFT
action.

It is known (see \cite{FN1} and \cite{FN2}) that at least
one instanton solution
exists for the 8 dimensional equation
$F^{\mu\nu}=\frac{1}{2}T^{\mu\nu\rho\sigma}
F_{\rho\sigma}.${\footnote {It is also known that a solution exists in
seven dimensions
if one replaces $Spin(7)$ by $G_2$ (see \cite{GN}).}\raisebox{1ex}{,}{\footnote
{An interesting problem is to find conditions on a curved compact 
Joyce manifold $M_8$ so that such instantons exist.}} 
For this solution, one has the important relation that
$\trace F_{[\mu\nu}F_{\rho\sigma]}$ is proportional to the
tensor $T_{  \mu\nu\rho\sigma}$. This property and the fact
that
$T_{\mu\nu\rho\sigma}$ is self-dual, will allow us later to
couple the pure 8D
Yang-Mills theory to a 3-form gauge field. Finally,
eqs.~(\ref{ims1})-(\ref{ims3}) imply
\beq
{4}\sum_{i=1}^7 \trace (\Phi_i \Phi_i)\cdot~(vol)~=~-\Omega
\wedge \trace(F \wedge F) +\trace (F  \wedge * F)~.
\label{squareJ}  
\eeq

\subsubsection{Action and observables} 

In the following all the fields are Lie
algebra valued and we will suppress the Lie algebra indices.
We use the standard
notation $(\psi_\mu, \phi)$ for topological ghost. We also
introduce the
Faddeev-Popov ghost $c$ to define a completely nilpotent
BRST transformation. The
topological BRST transformation for the gauge field and the
ghost fields is
\beqa
{s} A_\mu &=& \psi_\mu + D_\mu c~, \quad
{s} \psi_\mu~=~-D_\mu \phi - [ c, \psi_\mu ]~, \CR
{s} c &=& \phi - \frac{1}{2} [c , c]~, \quad
{s} \phi~=~-[ c, \phi ]~.
\eeqa
We need as many pairs of the anti-ghost and the auxiliary
fields $(\chi_i,
H_i)$ as topological gauge functions,
with the following BRST transformation law;
\beq
{s} \chi_i = H_i - [c, \chi_i]~, \quad
{s} H_i = [\phi, \chi_i] - [c, H_i]~.
\eeq
One has $1\leq i\leq 7$. The gauge fixed action at the first
stage is
\beqa
S_1&=&~\frac{1}{2} \int_{M_8}~ \Omega \wedge
\trace ( F \wedge F)~ + {s} \biggl[ \frac{1}{2} \int_{M_8}
d^8 x
\sqrt{g}~\trace (\chi_i \Phi_i + \frac{1}{2} \chi_i H_i)
\biggr]  \CR
&=&  ~\frac{1}{2} \int_{M_8}~ \Omega \wedge \trace ( F
\wedge F)~\CR
 && + \frac{1}{2} \int_{M_8} d^8 x \sqrt{g}~\trace
\biggl(H_i \Phi_i
+ \frac{1}{2} H_i H_i - \chi_i  (D\psi)_i + \frac{1}{2} \phi
[\chi_i, \chi_i]
\biggr)~,
\eeqa
where $(D\psi)_i$ is the FP ghost independent part of ${s}
\Phi_i$.  Eliminating the auxiliary fields $H_i$ by eq.~(\ref{squareJ}),
one  recovers the standard Yang-Mills kinetic term
\beq
S_1~=~ \int_{M_8} d^8 x \sqrt{g}~\trace \biggl( -\frac{1}{4}
F^{\mu\nu}
F_{\mu\nu} - \chi_i (D\psi)_i + \frac{1}{2} \phi [\chi_i,
\chi_i] \biggr)~.
\eeq

The ordinary gauge fixing and  Faddeev-Popov ghost
dependence have not been
considered yet: the first stage action has still a gauge
symmetry
in the ordinary sense. To fix it completely we take two more
conditions;
\beq
D\cdot\psi ~=~0~, \quad \partial\cdot A ~=~0~.
\eeq
(The meaning of the scalar product is the usual one, e.g.
$D\cdot \Psi=D_\mu \Psi^\mu$.)
Introducing additional fields $(\bar\phi, \eta)$ and $(\bar
c, B)$
with the BRST
transformation law;
\beqa
{s} \bar\phi &=& \eta - [c, \bar\phi]~, \quad {s} \eta~=~ [
\phi, \bar\phi]-
[c, \eta]~, \CR
{s} \bar c &=& B - [c, \bar c]~, \quad {s} B ~=~ [ \phi, \bar
c] - [c, B]~,
\eeqa
we write the complete action as
\beqa
S_2 &=& S_1+ {s} \biggl[  \int_{M_8} d^8 x \sqrt{g}~\trace
(\bar\phi
D\cdot\psi + \bar c \del\cdot A  + \frac{1}{2} \bar c B)
\biggr]  \CR
&=& \int_{M_8} d^8 x \sqrt{g}~\trace \biggl[ -\frac{1}{4}
F^{\mu\nu}
F_{\mu\nu}  - \chi_i (D\psi)_i + \frac{1}{2} \phi
[\chi_i, \chi_i] \CR
& & +\eta D\cdot \psi + \bar\phi D\cdot D \phi -\psi
\cdot[\bar\phi, \psi ] + B \del \cdot A + \frac{1}{2} B^2
+ \bar c \del\cdot D c \CR
& & - \bar c \del\cdot \psi  + \del\cdot A [c, \bar c]
- \frac{1}{2} \phi [\bar
c, \bar c] \biggr]~.
\label{action}
\eeqa

 A natural set of topological observables is derived from
the topological
invariants
\beq
\frac{1}{2} \int_{M_8} \Omega \wedge \trace (F \wedge F)~,
\quad
\int_{M_8} \trace (F \wedge F \wedge F \wedge F)~.
\eeq
The method of the descent equation implies a ladder of
topological invariants
and, for example, gives the following descendants;
\beqa {\cal O}^{(0)} &=& \frac{1}{2} \int_{M_8} \Omega
\wedge \trace (F \wedge F)~,\CR
{\cal O}^{(1)} &=& \int_{\gamma_7} \Omega \wedge
\trace ( \psi \wedge F)~, \CR
{\cal O}^{(2)} &=& \int_{\gamma_6} \Omega \wedge
\trace ( \frac{1}{2} \psi \wedge
\psi - \phi \wedge  F)~, \label{obs} \\
{\cal O}^{(3)} &=& - \int_{\gamma_5} \Omega \wedge
\trace (\psi \wedge \phi)~, \CR
{\cal O}^{(4)} &=& \frac{1}{2} \int_{\gamma_4}
\Omega \wedge  \trace ( \phi \wedge \phi)~. \nonumber
\eeqa
The descendant ${\cal O}^{(k)}$ with ghost number $k$ is an
integral over an
$(8-k)$ cycle $\gamma_{(8-k)}$.

\subsubsection { Geometric interpretation}

The virtual dimension of the moduli 
space ${\LM _J}$ of solutions to $P_+ F_A =0$ is $-index \
\dsla \otimes I_ {\LG}$,
i.e., the index of $\dsla \otimes I_ {\LG}: S^-\tensor{\LG }\to S^+\tensor
{\LG}$.
Its value is
\beq
-\int_{M_8} \hat A(M_8)~{\rm ch}(\LG)~,
\eeq
computable in terms of the relevant characteristic classes. 
We will discuss the vanishing theorem needed to make
the virtual dimension equal to the actual dimension
elsewhere.

We can interpret section 2.1.1 geometrically analogous to
section 5 in
\cite{BS}.
The BRST equations in this section are the analogues of (7)
in \cite{BS}, and are
the structure equations for the universal connection on
${\LA/ \LG} \times M_8$ with
structure group $G$.  The curvature 2-form ${\LF}$ for this
universal
connection equals ${\LF}_2^0 +{\LF}_1^1+{\LF}_0^2$,
where ${\LF}_{2-i}^i $ is an $i$-form in the
$\LA/ \LG$ direction (ghost number) and a $(2-i)$-form
in the $M_8$ direction. Note
that ${\LF}^{0}_2 $ at $(A,x)$ is $F_A(x)$ and ${\LF}_{1}^1
$ assigns to $\tau
 \in T({\LA/ \LG} ,A)$ and $v \in T(M_8,v)$ the value
$\tau(v)\in {\LG}$, since
$\tau$ is a 1-form on $M_8$.
Further, ${\LF}_{0}^2$ on $\tau_1,\tau_2 \in
T({\LM}_J, A)$ is $G(b^*_{\tau_1}(\tau_2))$ where
$G=(D_A^*D_A)^{-1}$ on $\L^0\tensor\LG$ and $b
_{\tau_1}(f)=[\tau_1,f]$
for $f \in \L^0\tensor\LG$; $b^*_{\tau_1}$ is the adjoint of
$b _{\tau_1}$.
We restrict $\LF$ to $\LM_J\times M_8$ and consider
$c_2={1\over{8\pi ^2}}\trace (\LF\wedge\LF)$ a 4-form
on $\LM_J\times M_8$. Its expansion contains
$  {1\over{8\pi^2}}\trace (\LF^1_1\wedge\LF^1_1)$,
which has ghost number 2. This 4-form assigns to $\tau_1,
\tau_2 \in
T(\LM_J , A)$ and $v_1, v_2 \in T( M_8, x)$
the value $ {1\over{8\pi^2}}(\trace
(\tau_1(v_1)\tau_2(v_2))- \trace (\tau_1(v_2)\tau_2(v_1))$.
Let $\tau_1{\tilde
\wedge}\tau_2$ denote this 2-form on $M_8$.

Let $c^{k}_{4-k}$ be the component of $c_2$ which is of
degree $k$
in the $\LM_J$ direction and of degree $4-k$ in the $M_8$
direction.
Then $\int_{\gamma_k }\O\wedge
c^{k}_{4-k}$ gives a $k$-form on $\LM_J$, when ${\gamma_k }$
is a
$(8-k)$-cycle on
$M_8$, $k=0 ,1,2,3$ or $4$. These are the observables ${\cal
O}^{(k)}$ in
eq.~(\ref{obs}).
Taking products of the forms $\LO$ and integrating them over
$\LM_J$ gives the
expectation values of the products of observables. {\it We are
not addressing the
central problem of integrating a form over the non compact
space} $\LM_J.$
We can specialize to 6-cycles, or equivalently to 2-forms
to get a closer analogy
to Donaldson invariants: if $\sigma \in H^2(M_8)$, let
$\Sigma_\sigma$ be the
2-form on $\LM_J$ given by $\Sigma_\sigma(\tau_1,\tau_2)=
\int_{M_8} \Omega\wedge\tau_1\tilde\wedge\tau_2\wedge\sigma$.
We get an $r$-symmetric multi-linear
function on $ H^2(M_8)$ given by $( \sigma_1, \ldots,
\sigma_r) \to \int_{\LM_J} \Sigma
_{\sigma_1}\wedge\ldots\wedge \Sigma _{\sigma_r} $, 
if ${\rm dim }~{\LM_J}=2r$.
Of course the issue here is to make these invariants
well-defined and to see how they depend on the space of
Joyce manifolds
modulo diffeomorphisms for a fixed $M_8$.


\subsection {Type H: Calabi-Yau Complex 4-manifold}

\subsubsection { Geometrical setup}

Suppose now that the holonomy group for ($M_8,g$) with
metric $g$ is $SU(4)$.
So $M_8$ is a complex manifold and we can assume
that $g$ is a Calabi-Yau metric with a
K\"ahler 2-form $\omega$. We choose a holomorphic
covariantly
constant (4,0)-form $\O$ which  trivializes the canonical
bundle $K$.
We normalize $\O$ so that
$\O \wedge \overline{\O}$ is the volume element of $M_8$.
We also choose the trivial $\sqrt K$ for the spin structure
on $M_8$.

We know that complex spinors can be identified with forms:
$S_M^\pm\otimes {\bf C}\simeq \L^{0, {{\rm even}\atop {\rm odd}}}$;
and the Dirac operator with $\bar\partial +\bar \partial^*$. Real
Majorana
spinors $S_M\subset S_M\otimes {\bf C}$ are the fixed points of
a conjugation $b$ on $S_M\otimes {\bf C}$.
We can identify $b$ with a conjugate linear $*$  operator as
follows.
For any Calabi-Yau $M_{2n}$, define $*: \L^{0,p} \to
\L^{0,n-p}$ by $\langle \alpha,\beta\rangle
=\int_{M_{2n}}\O
\wedge\alpha\wedge \ *\beta$, where now $\O\in \L^{n,0}$.
(If one denotes by $*_1$ the usual map on complex manifolds:
$\L^{p,q} \to \L^{n-q,n-p}$, then $*_1\ ^-=\O\wedge*$ on
$\L^{0,q}$.)
When $n=4$, one can show
that conjugation $b$ equals $(-1)^q*$ on $\L^{0,q}$.
Consequently, the operator $\bar
\partial^*+P_+\bar\partial: \,  \L^{0,1}\to
\L^{0,0}+\L^{0,2}_+$ is the Dirac
operator from
$S_M^- \to S_M^+$. Here $\L^{0,2}_\pm $ is the $\pm$
 eigenspace of $*$, $P_\pm$ is the projection of
$\L^{0,2}$ on $\L^{0,2}_\pm;$
we have identified $\L^{0,1}$ with $\frac{1-*}{2} (\L^{0,1} 
+ \L^{0,3})$ and $\L^{0,0}$ with ${{{\bf 1}+*}\over
2}(\L^{0,0}+\L^{0,4})$. The sequence $\L^0 {\stackrel{\bar
\partial}{\longrightarrow}}\L^{0,1} {\stackrel{P_+\bar
\partial}{\longrightarrow}}\L^{0,2}_+$ is elliptic and is
 the linearization of the equation $P_+F_A=0,$ modulo gauge transformations.

Suppose now $(E,\rho)$ is a complex Hermitian vector bundle
over
$M_8$ with metric $\rho$ of
${\rm dim}_{\bf C} = N$. If $A$ is a connection for $E$, we
have its covariant
differential $D_A: C^\infty(E)\to C^\infty (E\otimes
\L^1)$ so that
$D_A=\partial_A+\bar\partial_A$ with
$\bar \partial_A :\ C^\infty (E)\to C^\infty
(E\otimes\L^{0,1})$.
By introducing local complex coordinates $z^\mu$
$\bar \partial_A
(\ss^I)= (\partial_{\bar \mu}+ (A^I_J)_{\bar \mu}\ss^J)
d\bar z^\mu$,
$I,J=1,\ldots,N$. So $(A^I_J)_{\bar \mu}d\bar z^\mu$ is
a (0,1)-form on $M_8$ with $N\times N$ matrix  coefficients.

The 1-form connection $A$ with values in $GL(N,{\bf C})$ does not
split naturally
into $\L^{0,1}+\L^{1,0}$ unless $E$ is holomorphic.
A splitting can be obtained by a
choice of almost complex structure on the principal bundle.
See Bartolomeis and Tian \cite{BarTian}. In any case, the curvature 
$F_A$ can be decomposed as
$F_A=F_A^{2,0}+F_A^{1,1}+F_A^{0,2}$ with
$F_A^{0,2}=\bar\partial_A^2$.

For each $\bar\partial$ operator: $C^\infty(E)\to C^\infty
(E\otimes \L^{0,1})$,
there exists a unique connection $A$ such that (i) $A$
preserves the hermitian
metric $\rho$ of $E$ and (ii)
$(D_A)^{0,1}=\bar \partial$. Hence, the space $\LA_P$ of
$\bar\partial$ operators can be identified with the
connections of the
principal
bundle $P$ associated with
$E$, which preserve the Hermitian metric. The group of
complex gauge
transformations $\LH$ acts on the space $\LA_P$, because if
$h \in \LH$,
then $ h^{-1} \bar\partial h$ is also a
$\bar\partial$ operator.

Let $\LG$ be $gl(N,{\bf C})$. Then the sequence
$\L^0 \otimes\LG {\stackrel{\bar \partial
_A}{\longrightarrow}} \L^{0,1} \otimes \LG {\stackrel{
P_+\bar \partial_A
}{\longrightarrow}}\L^{0,2}_+ \otimes\LG $ is still elliptic
on the symbol
level.
We say $\bar\partial_A$ is holomorphic anti-self-dual if
$P_+F_A^{0,2}=0$,
in which case the sequence is elliptic.
Its index is the index of $\dsla \otimes I_{\LG }: S_M^-
\otimes
\LG \to S_M^+\otimes \LG$.

Again, the BRSTQFT will be obtained by gauge fixing
$S_0=\int_{M_8}\O\wedge \trace
(F_A^{0,2}\wedge F_A^{0,2}).$ 
 $S_0$ is independent of $A$, because $S_0= 8\pi^2 \Omega \cup p_1(E)$, 
since $\Omega \in \Lambda^{4,0}.$ 
When $S_0\neq 0$, we can normalize $\O$ further
by $e ^{i\theta}$, so that $S_0$ is real and positive.

To verify eq.~(\ref{square}) in the H case, we reduce $\LG$ to $u(N),$ 
using the metric $\rho.$  If $\o \in
\L^{0,2}$ has
components $\o_{\pm}$ in $\L_\pm^{0,2}$, then
$\parallel\!\o\!\parallel^2 =\parallel\!\o_+\!\parallel^2+\parallel\!\o_-\!
\parallel^2$.
And
\beqa
 0\leq S_0&=&\trace \int_{M_8}
\O\wedge(F^{0,2}_{A+}+F^{0,2}_{A-
})\wedge(F^{0,2}_{A+}+F^{0,2}_{A-})\CR
&=&-\parallel\! F^{0,2}_{A+}\!\parallel^2+\parallel\!
F^{0,2}_{A-}\!\parallel^2+i
{\rm Im}~\langle F^{0,2}_{A+} ,F^{0,2}_{A-}\rangle \CR
&=&-\parallel\! F^{0,2}_{A+}\!\parallel^2 +\parallel\!
F^{0,2}_{A-}\!\parallel^2~.
\eeqa
Hence
\beqa
 \parallel\! F^{0,2}_{A}\!\parallel^2=2 \parallel\! F^{0,2}_{A
+}\!\parallel^2 +S_0~.
\eeqa
So the holomorphic
anti-self-dual gauge condition minimizes the action
$\parallel\! F^{0,2}_{A}\!\parallel^2$ in the
topological sector with $S_0$ fixed.

The $(4,0)$ form $\Omega$ can be simply expressed in local coordinates as
\beq
\Omega= dz^1\wedge dz^2 \wedge dz^3 \wedge dz^4~.
\eeq
And
\beq
F^{(0,2)}=d\z^{\bar\m} d\z^{\bar\n} F_{\bar\m\bar\n}~,
\eeq
where
\beq
F_{\bar\m\bar\n}=\pa_{\bar\m} A _{\bar\n}-\pa_{\bar\n} A
_{\bar\m} +[ A
_{\bar\m}, A _{\bar\n}]~.
\eeq
also
\beq
D_{\bar\m}=\pa_{\bar\m} +[ A _{\bar\m},~~~]~.
\eeq
One has the part of the Bianchi identity
\beq
 D_{[ \bar \mu }
F_{\bar \nu\bar \rho ]}=0~.
\eeq

The 3 complex gauge covariant gauge conditions,
which count for 6 real conditions
on the 8 independent real components contained
in $A_{\bar \m}$ are
\beq
F_{ \bar \m_1\bar \m_2}+\epsilon_{\bar \m_1 \bar \m_2\bar
\m_3\bar \m_4} F_{
\bar \m_3\bar \m_4} = 0~.\label{ct}
\eeq The complex Landau type condition is
\beq
\pa _{ \bar \m} {^{\bf c } A_{ \bar \m}} = 0~.\label{cl}
\eeq
We have now the topological ghost $\Psi_{ \bar \m}$ with 4
independent
complex components, and we have the ghost gauge condition
\beq
 D _{ \bar \m} \Psi_{ \bar \m} = 0~.
\eeq
(Here and below, we    use the left upper symbol $^{\bf c}$ for complex
conjugation.) A consequence of the use of
complex gauge transformations is that a complex Faddeev-Popov ghost $c$
must   be introduced, with complex ghost of ghost $\phi$.
Up to the complexification of
all fields, we have thus exactly the same field content as
the original 4 dimensional
Yang-Mills TQFT. This  leads us to the BRST algebra that we will
shortly display.

\subsubsection{Action and observables}

~From the previous arguments, we must write
the BRST algebra in a notation where all fields are complex
fields and replace the formula of the
J  case by
\beqa
s A_{\bar \mu} &=& \psi_{\bar \mu} + D_{\bar \mu} c~, \quad
s \psi_{\bar\mu}~=~-D_{\bar \mu} \phi - [ c, \psi_{\bar \mu} ]~, \CR
s c &=& \phi - \frac{1}{2} [c , c]~, \quad s \phi~=~-[ c,
\phi ]~.
\eeqa
and in the antighost sector ($i$ runs between 1 and 3)
\beq
s \chi_i = H_i -[c, \chi_i]~, \quad s H_i =  [\phi, \chi_i]
- [c, H_i]~.
\eeq
Using the 3+1 complex gauge conditions eqs.~(\ref{ct}) and
(\ref{cl}), we get
\beqa
Z= \int & &[\D A_{ \bar \m}][\D ^{\bf c} A_{ \bar \m}][\D \Psi_{
\bar \m}] [\D
^{\bf c}\Psi_{ \bar \m}][\D \kappa_{ \bar \m \bar \nu}] [\D
^{\bf c}\kappa_{ \bar \m \bar
\nu}][\D H_{ \bar \m \bar \nu}] [\D ^{\bf c} H_{ \bar \m \bar
\nu} ]\CR& &[\D
 \eta ]
[\D ^{\bf c}\eta ][\D \phi ] [\D ^{\bf c}\phi ][\D \bar\phi ] [\D
^{\bf c}\bar\phi ][\D c
] [\D
^{\bf c} c ][\D \bar c] [\D ^{\bf c}\bar c][\D B] [\D ^{\bf c} B]\CR
& & \quad \exp \int [\Omega
\wedge \trace F^{(0,2)} \wedge F^{(0,2)} ] \CR
& & \quad  \exp
\int d^4 z d^4\z\ {s} \biggl[ \trace \biggl(\ \kappa_{ \bar
\m \bar \n}
( F_{ \bar\m \bar \n}+\epsilon_{ \bar \m \bar \n \bar \rho
\bar \sigma}
F_{ \bar \rho \bar\sigma}+{1\over 2} H_{ \bar \m \bar \nu})
\CR
& & \quad \quad \quad \quad
+{^{\bf c}\kappa}_{ \bar \m \bar \n}( ^{\bf c}F_{ \bar \m \bar \n}+
\epsilon_{ \bar \m \bar \n
\bar \rho \bar \sigma}{^{\bf c}F_{ \bar \rho \bar \sigma}
+{1\over 2} {^{\bf c} H}_{ \bar \m
\bar \nu} )}+\bar\phi D_{ \bar \m } \Psi_{ \bar \m}+ ^{\bf c}
\bar\phi  ^{\bf c}D_{
\bar \m }
^{\bf c}\Psi_{ \bar \m}\CR
 & &\quad \quad \quad \quad +\bar c (\pa _{ \bar \m} {^{\bf c}A}_{
\bar \m}
+{1\over 2} {^{\bf c} B})+^{\bf c}
\bar c ({^{\bf c}\pa }_{ \bar \m}{ A}_{ \bar \m}+{1\over 2} { 
B})\biggr)
\biggr] ~.
\eeqa
Eliminating the auxiliary fields, we have
\beqa
\label{partsym}
Z=\int & &[\D A_{ \bar \m}][\D ^{\bf c} A_{ \bar \m}][\D \Psi_{
\bar \m}] [\D
^{\bf c}\Psi_{ \bar \m}][\D \kappa_{ \bar \m \bar \n}] [\D
^{\bf c}\kappa_{ \bar \m
\bar \n}]
\CR
& &[\D \eta
 ] [\D ^{\bf c}\eta ] [\D \phi ] [\D ^{\bf c}\phi
][\D \bar\phi ] [\D
^{\bf c}\bar\phi ] [\D c ] [\D ^{\bf c} c ][\D \bar c] [\D ^{\bf c}\bar
c ] \CR
& &
\quad \quad  \exp\int_{M_8} d^8 x \sqrt{g}~\trace \biggl[-
\frac{1}{4}
F_{\bar\mu\bar\nu} {^{\bf c}}F_{\bar\mu\bar\nu} -
\chi_{\bar\mu\bar\nu} D_{\bar\mu }\psi_{\bar\nu }+
\frac{1}{2} \phi
[\chi_{\bar\mu\bar\nu}, \chi_{\bar\mu\bar\nu}]
\CR
& & +\eta
D_{\bar\mu } \psi_{\bar\mu } + \bar\phi D_{\bar\mu } D
_{\bar\mu}\phi
+\bar\phi[\psi  _{\bar\mu }, \psi _{\bar\mu }]+ B
\del_{\bar\mu } {^{\bf c} A}_{\bar\mu } +
\frac{1}{2} B  {^{\bf c}B }+ \bar c \del_{\bar\mu }D _{\bar\mu } c \CR
& & - \bar c
\del_{\bar\mu } \psi_{\bar\mu } + \del_{\bar\mu } A_{\bar\mu
} [c, \bar c] -
\frac{1}{2} \phi [\bar c, \bar c]\  +\ {\rm complex \
conjugate}\ \biggr]~.
\eeqa
We can replace $ d^8x F_{\bar\mu\bar\nu}
{^{\bf c}}F_{\bar\mu\bar\nu}$ by
$ \trace (F\wedge *F)$, because 
$\frac{1}{4} \parallel\! F_A \!\parallel^2 =~
\parallel\! F^{0,2}_A \!\parallel^2 + 
\frac{1}{4} \parallel\! \langle F, \omega \rangle \!\parallel^2 + $
topological terms \footnote{The second term in the right hand side of this 
equation is really a gauge fixing term, as explained in section 2.3.}.  
See \cite{BarTian}, Proposition 3.1.

 The definition of observables follows
from
the cocycles obtained by the descent equations,
as sketched in the previous section. Their meaning is now
discussed.

\subsubsection{Geometric interpretation}

Let $\widetilde \LM$ denote $[A\in \LA_p]$ with
$F_+^{0,2}=0$.
It is invariant under
$\LH$ (which acts on $\LG$ in $\L^2_+\otimes \LG$, but not
on $\L^2_+$.)  Let
$\LM_H= \widetilde \LM /\LH$.
The 3 complex covariant gauge conditions, $\  \L^{0,2}_+=0$,
probe the moduli space $\LM _H$. We remarked earlier that
$0\to \L^0 \otimes\LG {\stackrel{\bar
\partial_A}{\longrightarrow}}
\L^{0,1} \otimes \LG {\stackrel{P_+\bar \partial _A
}{\longrightarrow}}
\L^{0,2}_+ \otimes\LG $ 
is an elliptic complex with $\bar \partial _A^*+P_+\bar
\partial _A : \L^{0,1}
\otimes \LG\to \L^{0} \otimes \LG + \L^{0,2}_+ \otimes \LG$;
the elliptic operator $\dsla_A:S^-\otimes\LG\to
S^+\otimes\LG$.
The {\it complex} gauge condition is $\bar\partial^* \tau=0$
for $\tau \in
\L^{0,1} \otimes\LG$.

As before, we get a hermitian vector bundle $\widetilde E$
over
$\LM_H\times M_8$ with connection. One can compute $c_2$ of
$\widetilde E$
in terms of its curvature $\LF^H$.  One has the map, $T$, of
$H^{0,*}(M_8)$ into forms on $\LM_H$ by $\mu
{\stackrel{T}{\longrightarrow}}\int_{M_{8}}\O\wedge \trace
(\LF^H\wedge
\LF^H)\wedge\mu$.

Formally this gives a multilinear map of $H^{0,*}(M_8)\to {\bf
C} $ by
$\mu_1,\ldots,\mu_r \to \int_{\LM_H} T(\mu_1)\wedge
\ldots\wedge T(\mu_r).$ These would be the expectation values
of
the observables of the BRSTQFT.

As in section 2.1.3., part of $c_2$ is $\frac{1}{8\pi^2} 
\trace ((\LF^H)^1_1 \wedge (\LF^H)^1_1)$ with 
$(\LF^H)^1_1 (\tau,v) = \tau(v)\in u(N).$  
If $\sigma\in H^{0,2}(M_8)$ let $T_\sigma$ be the 2-form on
$\LM_H$ given by
$\int_{M_{8}}\O\wedge \trace (\tau_1\wedge
\tau_2)\wedge\sigma $
where $\tau_i$, $i=1,2$, are (0,1) forms on $M_8$ with values in $u(N).$
The formal holomorphic Donaldson polynomial is the symmetric
$r$-multilinear function
on $H^{0,2}{(M_8)}$ given by $\sigma_1,\ldots,\sigma_r \to
\int_{\LM_{H}}
T_{\sigma_1}\wedge\ldots\wedge T_{\sigma_r}$, when ${\rm dim}~
\LM_H =2r$.
(Note that if $H^{0,2}(M_8)\neq 0$, then $M_8$ is
hyperK\"ahler
because elements of $H^{0,*}$ are covariantly  constant.)

It will be very interesting to see when formal integration
over $\LM_H$ is
justified, and when these invariants depend only on the
complex structure of
$M_8$, not on the Calabi-Yau metric $g$, nor the hermitian
metric $\rho$.
 C.~Lewis [12] is investigating the conditions under which
$\LM_H$ is
the set of stable holomorphic vector bundles.

Since the elliptic operator here is $\dsla$ again, the
virtual dimension of
$\LM_H$ is 
 \beq
  - \int _{M_{8}}\hat{A}(M_8) \ {\rm ch} \ (\LG)~.
\eeq


 \subsection {Comparison of H and J cases}

Under suitable conditions ($(E,\rho)$ a stable\footnote{For physicists, 
one might define $(E,\rho)$ to be stable if it is holomorphic, 
Einstein-Hermitian, i.e., $F_\rho \cdot \omega$ is a constant 
multiple of the identity, 
where $F_{\rho}$ is the curvature of $(E,\rho)$ relative to 
its unique $\rho$-connection.} vector bundle, for example),  one expects that 
the orbit space of $\LA_P$ under the group of complex gauge transformations, 
will be the same as the sympletic quotient, $\LA_P\parallel \LG_U,$  
where $\LG_U$ are the gauge transformations on $P$ reduced to the compact group 
$U(N).$  Since $[A; \langle F^{1,1}_A, \omega \rangle_m=0, \ m \in M_8 ]$ is 
the zeros of the moment map, 
$\LA_P/\LG_U$ is the orbit space of this set under $\LG_U.$

We replace the condition $P_+(F^{0,2}_A)=0$ with $F^{0,2}_A \in 
\Lambda^{0,2}_+ \otimes gl (N, {\bf C})$ by the conditions
$P_+(F^{0,2}_A)=0$ and
 $\langle F^{1,1}_A, \omega \rangle = 0$, where now $F^{0,2}_A \in
\Lambda^{0,2} 
\otimes u(N)$ 
and $\langle F^{1,1}_A, \omega \rangle  \in u(N).$ 
One should get the same moduli space of solutions.

In the linearization, the sequence 
$gl(N, {\bf C}){\stackrel{\bar \partial _A}{\longrightarrow}} \L^{0,1}
\otimes gl (N, {\bf C}){\stackrel{P_+\bar
\partial_A}{\longrightarrow}}\L^{0,2}_+ \otimes gl(N, {\bf C})\rightarrow 0$
is replaced by $u(N) \rightarrow \Lambda^{0,1} \otimes u(N)
 {\stackrel{P_{+}\bar\partial_{A} \oplus i_{\omega}\partial}{\longrightarrow}} 
\Lambda^{0,2}_+ \otimes u(N) \oplus u(N) \rightarrow 0$,
 where $i_\omega \partial: 
\tau \in \Lambda^{0,1} \otimes u(N) \rightarrow 
\langle \partial \tau, \omega \rangle_{m} \in u(N).$  
The operator $i_\omega \partial$ is the linearization of 
the $0$-momentum condition 
$\langle F^{1,1}_A, \omega \rangle _{m} = 0;$ it is also the 
imaginary part of $\bar{\partial}^*_A: \Lambda^{0,1} 
\otimes gl(N, {\bf C}) \rightarrow gl(N, {\bf C})$.
Thus, with the reduction of $Spin(7)$ holonomy to $Spin(6) = SU (4)$ holonomy, 
the 7~dimensional 
$\Lambda^2_+ $ in the J-case decomposes into the 
6 dimensional $\Lambda^2_+ $ of 
the H-case plus ~{\bf R}.


\subsection{Link to twisted supersymmetry}

We note that the field content of our Yang-Mills
BRSTQFT action in 8
dimensions is similar to that of four dimensional
topological Yang-Mills theory.
Since four dimensional topological Yang-Mills theory is a
twisted version of
$D\!=\!4,~N\!=\!2$ super Yang-Mills theory and  is related by
dimensional reduction to the
minimal six dimensional supersymmetric Yang-Mills theory, it is
natural to expect a
similar
connection in eight dimensions. This is indeed so; we
explain the type J case,
 although the
fields $(c, \bar c, B)$ which were employed to impose the
Lorentz condition
$\del^\mu A_\mu =0$, are neglected. The gauge
supermultiplet in eight
dimensions consists of one gauge field in ${\bf 8_v}$ (the
vector
representation),
one chiral spinor in ${\bf 8_s}$, one anti-chiral spinor in
${\bf 8_c}$ and two
scalars \cite{SS}. The reduction of the holonomy group to
$Spin(7)$ defines
decomposition of the chiral spinor; ${\bf 8_s} = {\bf 1}
\oplus {\bf 7}$.
Now it is
natural to identify $A_\mu$ and $\psi_\mu$ in our
topological theory as ${\bf
8_v}$ and ${\bf 8_c}$, respectively. Furthermore $\chi_i$
and $\eta$ just
correspond to the chiral spinor ${\bf 8_s}$ according to the
above
decomposition.
Finally $\phi$ and $\bar\phi$ give the remaining two
scalars. This exhausts
all the
dynamical fields in our action of eight dimensional
topological Yang-Mills
theory.
Though we do not work out the transformation law explicitly,
we believe
this is a
sufficiently convincing argument for the fact that the J case is a twist of
the  
$D\!=\!8$ SSYM dimensionally reduced from $D\!=\!10,~N\!=\!1$ SSYM.  The
connection between  a general supersymmetry transformations and 
  topological  BRST transformations  is the following:
when  $M_8$ is flat, the reduction from 
$D\!=\!10$ is $N\!=\!2$ real supersymmetry or  
$N\!=\!1$ complex supersymmetry. For curved manifolds, 
the only surviving supersymmetries  are those depending on 
    covariant constant
spinors. In the $J$ case the 
 nilpotent topological BRST symmetry generator 
 is a  combination  of the real and imaginary
parts of the  one surviving complex generator of  supersymmetry.

As said just above, this supersymmetric Yang-Mills theory in eight 
dimensions is
obtained by
dimensional
reduction from the $D\!=\!10$, $N\!=\!1$ super Yang-Mills 
theory. This  suggests a relationship  with superstring theory.
It has been
argued that the effective world volume theory of the D-brane
is the dimensional
reduction of  the ten dimensional super Yang-Mills theory
\cite{Wi2}. Thus the
BRSTQFT constructed in this section may arise as an
effective action of 7-brane
theory. In fact Joyce manifolds are discussed in
connection with
supersymmetric
cycles in \cite{BSV} \cite{BMOY}. Recently in \cite{FKN}, a six
dimensional
topological field theory of ADHM sigma model is obtained as
a world  volume
theory
of D-5 branes. The world volume theory of D-branes could  provide a
variety of higher dimensional BRSTQFT's.


\renewcommand{\theequation}{3.\arabic{equation}}\setcounter
{equation}{0}

\section{Coupling  of the 8D theory to a 3-form}

For the pure Yang-Mills theory, we have seen that the
construction of a BRSTQFT
 implies a
consistent breaking of the
$SO(D)$ invariance. This turns out to be  quite natural,
when closed but not
exact forms exist, like the K\"ahler 2-form on K\"ahler manifolds or the
holomorphic
$(n,0)$-form on Calabi-Yau manifolds.

This idea extends to consider  BRSTQFT's involving sets of
possibly interacting
$p$-form gauge fields with  $(p+1)$-form curvatures
$G_{p+1}=dB_p+...$,
satisfying relevant Bianchi identities.
Our point of view is that
one must define a system of equations, eventually
interpreted in BRSTQFT as
 gauge conditions, which does not overconstrain the fields.
If    tensors  $T^{\mu_1,\ldots ,\mu_{2p+2} } $ of    rank $2p+2; (2p+2
\leq D)$ exist which are invariant under maximal
subgroups of $SO(D)$,  we can   consider BRSTQFT  based on gauge functions 
of the following type, where $\lambda$ is a parameter:
\beq
T^{\mu_1,\ldots , \mu_{2p+2 } } G_{\mu_{p+2},\ldots
,\mu_{2p+2 }}~
=~\lambda G^{\mu_{ 1},\ldots ,\mu_{p+1 }}~.
\label{instp}
\eeq
Such equations  must be understood in a matricial form,
since they
generally involve
several forms $B_p$, with different values of $p$.
To ensure that the problem is well defined, a  first
requirement is that
eq.~(\ref{instp}) has solutions in
$G_{p+1}$ for
$\lambda$ different from zero. This  algebraic question
is in principle
straightforward to solve by group theory arguments,
although we expect that geometrical arguments should also
justify them.
Moreover,
 we must also consider that $G_{p+1}$ is the
curvature of a $p$-form gauge field $B_p$.  Thus, other
gauge functions must
be introduced, to gauge fix the ordinary gauge freedom  of
$B_p$ which
leave invariant its curvature $G_{p+1}$. This gives a second
requirement, since
from the point of view of the quantization,
the total number of gauge conditions, the topological ones
and the ordinary
ones, must be exactly equal to the number of independent
components in the
gauge field $B_p$.

To be more precise, the number of
ordinary gauge freedom of a $p$-form gauge field in $D$ dimensions
 is $C_{D-1}^{p-1}$:
(this amounts to the fact that $B_p$ is truly defined up to
a $(p-1)$-form,
which
is itself defined up to a $(p-2)$-form, and so on.)  We
should  therefore only retain invariant tensors $T  $ such
that the number of components of $B_p$ equates the rank
of the
system of linear equations  in $G$ presented in
eq.~(\ref{instp}) plus the
number of ordinary
gauge freedom in $B_p$.
Obviously, when there are several   fields in
eq.~(\ref{instp}),
the counting of independent conditions can become quite
subtle, since one must
generally  combines
several equations like eq.~(\ref{instp}). For instance,
 we will   display  in the
next section BRSTQFT theories in dimensions $D<8$. Their
derivation will
appear as rather simple, because they all descend by
dimensional reduction
from the  pure Yang-Mills BRSTQFT   based on the
set of 6 or 7 independent self-duality gauge covariant
equations in eight
dimensions found in section 2. Without this insight, their derivation  would be
less obvious.

We now turn to the introduction of a 3-form gauge field in 8
dimensions.
In even $ D=2k$ dimensions, eq.~(\ref{instp})  has a generic
solution  for 
an uncharged   ~$(k-1)$-form~ gauge field $B_{k-1}$:
assuming the existence of a curvature $G_{k }$ for $B_{k-1}$, we can
consider the
obvious generalization of self-duality equations,
$G_{k }= *G _k$.
The number of these conditions is
$C^k_{D-1}$.
 On the other hand, the number of ordinary gauge freedom of
a ~($k-1$)-form~ gauge field is
$
C^{k-2}_{ D-1}=C^{k-2}_{ D}-C^{k-3}_{ D}+C^{k-4}_{ D}-
\ldots \pm C^{0}_{ D }~.
$
Thus imposing the ordinary gauge fixing  conditions for
 the ($k-1$)-form
gauge field
plus the gauge covariant ones, $G_{k }=*G _k$, gives a number of
$C^{k-1}_{ D} =
C^{k-2}_{ D-1}+C^k_{ D-1}$
equations,
which is equal to the number of arbitrary local
deformations of
the   $C^{k-1}_{ D}$
independent components of the  ($k-1$)-form gauge field.
We will see that it is possible to  generalize the    
   the self duality equation satisfied by  a 
($k-1$)-form gauge field.
Moreoever, the counting remains correct in the case it has a charge.
As an example, in the
8-dimensional theory,   a 3-form gauge field has 56
components,
with 21 ordinary gauge freedom, while
the number of self dual equations involving the  4-form
curvature
 of the 3-form is 35, and one has 56=21+35.

We thus propose as topological gauge conditions for the
coupled system
made of the Yang-Mills field $A$ and the 3-form gauge field
$B_3$ the following
{\it coupled} equations;
\beqa
\lambda  F_{\mu\nu}&=& T_{\mu\nu\rho\sigma}F_{\rho\sigma}~,
\CR
dB_3+ *(dB_3)&+&\alpha \trace (F\wedge F) =0~.
\label{truc}
\eeqa
 $\alpha$ is a real number, possibly quantized \footnote{ 
  Equation  (\ref{truc}) suggests that the 3-form could be involved 
in an anomaly compensating mechanism.}. Although
$B_3$ is real valued,
it interacts with the  Yang-Mills connection $A$, when  $\alpha\neq 0$.
 An octonionic instanton
 solves the first equation; for  the second equation to hold, 
$\trace (F\wedge F)$ must be self-dual.
A solution in $B_3$ is given by  eqs. (25), (30), (31) of
\cite{GN}, in the case of
$M_8=S^7\times {\bf R}$.

Given these facts, we are led to define a BRSTQFT  in 8 dimensions  based
on the gauge conditions (\ref{truc}),
in which 
a  3-form gauge field is coupled to a Yang-Mills field. 
 The ghost spectrum for the ordinary gauge invariance of the
field
$B_3$ generalizes
that of the Yang-Mills field, with the following unification
between the ghost
$B^1_2$ and  the ghosts of ghosts $B^2_1$ and $B^3_0$
\beq
\widehat B_3=
B_3+B^1_2+B^2_1+B^3_0~. \label{bghost}
\eeq
(From now on upper indices mean ghost number and lower
indices ordinary
form degree.)

The BRST symmetry of the topological Yang-Mills symmetry
considered in the
previous section satisfies
\beq
\widehat A= A+c~,
\eeq
\beq
\widehat F=
(s+d)\widehat A+{1\over 2}[\widehat A,\widehat
A]=F_2^0+\Psi^1_1+\phi^2_0~,
\label{ym}
\eeq
with the notation  $\Psi^1_1 =\Psi_\m dx^\m$ and $
 \phi^2_0=\phi $.

The gauge symmetry of the 3-form $B_3$ involves a 2-form
infinitesimal
parameter associated to $ B^1_2$.  We can distinguish
however different
topological sectors for $B_3$, which cannot be connected
only by these
infinitesimal gauge transformations.
 As an example, $B_3$ and $B'_3$ can
belong to such different sectors, if
\beq
  B'_3= B_3 + \trace (  A\wedge dA +{2\over3}A\wedge A\wedge
A)~,
\label{redef2}
\eeq
We thus define the curvature  of  $B_3$ as
\beq
  G^{(A)}_4=dB_3 +  \trace (  F^{(A)}\wedge F^{(A)})~,
\label{redef3}
\eeq
where the index ${(A)}$ means the dependence upon the Yang-
Mills field $A$.
 Notice that it is not globally possible to eliminate the
$A$ dependence of
 $G^{(A)}_4$ by a field redefinition of $B_3$ involving the
Chern-Simons
 3-form.

The topological BRST symmetry of the 3-form gauge field
 system is defined from
\def\Gi{{\cal G}}
\beqa
\widehat G _4 =(s+d)
\widehat B_3
+ \trace ( \widehat F^{(A)}\wedge \widehat F^{(A)})
=G _4+\Gi_3^1
+\Gi_2^2+\Gi_1^3+\Gi_0^4~,
\eeqa
that is
\beqa
(s+d)&(
B_3+B^1_2+B^2_1+B^3_0)+
\trace  \left((F^0_2+\Psi^1_1+\phi^2_0)\wedge
(F^0_2+\Psi^1_1+\phi^2_0)
\right)
=\CR  &d
B_3 +
\trace  ( F \wedge F) +\Gi_3^1
+\Gi_2^2+\Gi_1^3+\Gi_0^4  ~.
\label{b3}
\eeqa
The fields $\Gi^g_{4-g}$, $g=1,2,3,4$ are the topological
ghosts of $B_3$.
By expansion in ghost number, Eqs.~(\ref{ym}) and (\ref{b3})
 define a BRST operation $s$ which,
eventually, determines the equivariant cohomology of
arbitrary deformations
of the Yang-Mills field modulo ordinary gauge
transformations and
of the 3-form gauge field, modulo the    infinitesimal
gauge transformations,
$\delta B_3=d\epsilon_2,\
 \epsilon_2\sim \epsilon_2+ d\epsilon_1,\
\epsilon_1\sim \epsilon_1+ d\epsilon$.

There is a natural topological invariant candidate for the
classical part
of a BRSTQFT action,
\beq
I_{top}=\int \Big ( G^{(A)}_4\wedge G^{(A)}_4 +\Omega\wedge \trace
(F^{(A)} \wedge F^{(A)})  \Big)~.
\label{inst3}
\eeq
Its  gauge fixing is   a generalization of what we do in
 the pure Yang-Mills case.  The main
point is to find the gauge function in the topological
sector.
The existence of the octonionic instanton, together with
an associated moduli space (yet to be explored),
indicates that  eq.~(\ref{truc}) is   a good choice 
\footnote{Notice that  one could also consider a 7-dimensional theory,
which is  formally related to the BRSTQFT in 8 dimensions as the
3-dimensional Chern-Simons theory is related to the 4-dimensional
Yang-Mills TQFT action.}.

To enforce the gauge function eq.~(\ref{truc}), one must
introduce a self-dual
4-form antighost $\kappa_{\mu\nu\rho\sigma}$, and consider
the following
BRST exact action
\beq
S_3 =
\int d^8x\ {s}\left(
\kappa^{\mu\nu\rho\sigma}(\partial_{[\mu}B_{\nu\rho\sigma]}
+ \epsilon^{ \mu \nu\rho\sigma\alpha\beta\gamma\delta}
\partial_{[\alpha}B_{\beta\gamma\delta]}
+\trace F_{[\mu\nu}F_{ \rho\sigma]})\right)~.
\eeq

The  remaining conditions are  for the usual
 gauge invariances of forms, whether
they are classical or ghost fields. One can choose
the following
gauge fixing conditions for the longitudinal parts of
all ghosts and ghosts of ghosts  $\Gi^g_{4-g}$
\beqa
\partial^\mu \Gi^1_{3\mu\nu\rho} &=&a \Delta^1_{\nu\rho}~,
\CR
\partial^\mu \Gi^2_{2\mu\nu} &=& b\Delta^2_{\nu}, ~ \CR
\partial^\mu \Gi^3_{1\mu} &=&
 c\Delta^3~,
\label{tg}
\eeqa
One must also conventionally   gauge fix the longitudinal 
components of $B_{ 3\mu\nu\rho}$,  of the ghosts $B^1_{
2\mu\nu }$ and
and $B^2_{ 1\mu  }$, and of the antighosts.
   The presence in the r.h.s. of
eq.~(\ref{tg}) of
the cocycles  $\Delta^g_{3-g}$ stemming from the ghost
decomposition of
$\trace\widehat F \wedge \widehat F=
\trace(F+\Psi+\phi)\wedge (F+\Psi+\phi)$
is an interesting possibility. It  can  lead  to mass
effects in TQFT, when
 the ghost of ghost $\phi$ takes a given mean value,
depending on  the
 choice of the vacuum in the moduli space, which can be
adjusted by
 suitable choices of the parameters $a, b,c$.

All these   gauge conditions can be enforced in a BRST
invariant way,
 as explained e.g. in \cite{blb}.  The final
 result is  an action of the following type (including the
pure
 Yang-Mills part discussed in the previous sections)
\beqa
S=\int (\pa_\m B_{ \nu\rho\sigma} \pa^\m B^{ \nu\rho\sigma}
+\trace F^{ \mu\nu}  F_{  \mu\nu}
+\pa_\m B_{ \nu\rho\sigma}
\trace F^{ \mu\nu}  F^{  \rho\sigma}\CR
+{\rm supersymmetric \  terms})~.
\eeqa

 The observables are defined from all forms
$\widehat O^g_{8-g}$ occurring in the
ghost expansion of the 8-form
\beq
\widehat O_8= \widehat G_4\wedge  \widehat G_4 ~.
\eeq
Whether these supersymmetric terms, made of ghost
interactions are
linked to Poincar\'e supersymmetry is an interesting question.

\subsection {Mathematical Interpretation}

Fix an element of $H^4(M_8, {\bf Z})$ and let $h_4$ denote its
harmonic representative.  Let $\ss$ denote the affine space
of all closed 4-forms  which represent  this cohomology class.
Then $\ss~=~h_4+d\Lambda^3;$ strictly speaking
$\ss=h_4+d(\Lambda^3/$closed 3-forms) $=h_4+d\delta \Lambda^4.$
In any case a tangent vector to $\ss$ can be represented as
$dB_3$ with $B_3$ a 3-form.

There are other ways of describing $\ss.$ An
element of $\ss$ can be represented as a collection of 
3-forms $\{B_u\},$ for a collection of coordinate
neighborhoods $U$ covering $M_8,$ satisfying $B_u -
B_v=dw_{u,v}$ on $u\cap v.$  Thus $\{dB_u\}$ gives
a well-defined closed form on $M_8;$ to be an element of $\ss,$
this 4-form must be cohomologous to $h_4.$ In the earlier
part of this section, $dB_3$ means this element of $\ss$ when
$B_3$ is defined locally as $B_{3,u};$ or if $B_3$ is an
ordinary three form, $dB_3$ is really $h_4 + dB_3.$\footnote{The theory
 of gerbes \cite{Bry} gives a sheaf theoretic description for
 exhibiting integral cohomology classes, extending the notion 
of curvature field as an integral 2-cocycle.}

Next consider the elliptic complex $0 \rightarrow \Lambda^0
\rightarrow \Lambda^1 \rightarrow \Lambda^2 \rightarrow \cdots
\rightarrow \Lambda^4_+ \rightarrow 0$, where $\Lambda^4_\pm$
are the $\pm 1$ eigenspaces of the ordinary $\ast$ operator
on $M_8.$  Remember that in the J-case we also had $0
\rightarrow \Lambda^0 \rightarrow \Lambda^1 \rightarrow
\Lambda^2_+ \rightarrow 0$ with $\Lambda^2 = \Lambda^2_- \oplus
\Lambda^2_+$ of dimensions 21 and 7, respectively.  Consider
then $0 \rightarrow \Lambda^2_- \stackrel{d}{\longrightarrow}
\Lambda^3 \stackrel{d}{\longrightarrow} \Lambda^4_+
\rightarrow 0.$  We leave the reader to check that it is
elliptic.  (It does not suffice that the dimensions are 21,
56 and 35, respectively.)  The linearization of the problem
below involves $0 \rightarrow \Lambda^0 \otimes \LG
\stackrel{D_A}{ \longrightarrow }\Lambda^1 \otimes \LG
\stackrel{D_A}{\longrightarrow}  \Lambda^2_+ \otimes \LG
\rightarrow  0$ for connections, and $0 \rightarrow
\Lambda^2_- \stackrel{d}{\longrightarrow} \Lambda^3
\stackrel{d}{\longrightarrow} \Lambda^4_+ \rightarrow 0$ for
3-forms.

     An analogue of the anti-self-dual equations for the pair
$(A,G)$ with  a connection $A$ and $G \in \ss$ is
\beqa
&{\rm (a) } & ~~~F_A~=~ *~\O \wedge F_A~, \hspace*{10ex} ({\rm i.e.} \
P_+ F_A=0)\quad  \CR
&{\rm (b) }& ~~~(1+*)G~=~ - \alpha \trace (F_A \wedge F_A)~.
\label{trucm}
\eeqa
This equation  is       a mathematical interpretation of  (\ref{truc}).
Note that if a solution $A, G = h_4 + dB_3$ exists for (\ref{trucm}), then
$\trace (F_A \wedge F_A)$ is self-dual and hence harmonic.
Hence $(1+*) (h_4+dB_3)$ is harmonic.  Since $(1+*)h_4$ is
harmonic, so is $(1+*) dB_3.$  Hence $dB_3 = 0,$ and
$G=h_4.$  Note also that the sector $\ss,$ i.e. the element
chosen in $H^4(M_8, {\bf R})$ must have its self-dual part, a
multiple of the self-dual element $p_1(P).$

If we linearize (\ref{trucm}), we get for $\tau \in T(\LG)$ and
$B_3 \in T(\ss),$ the equations $P^{(2)}_+ \ (D_A\tau)=0$
and $P^{(4)}_+dB_3 = 0$, where $P^{j}_+$ is the
projection of $\Lambda^j \rightarrow \Lambda^j_+\quad (j=2,4) .$
We then have a pair of elliptic systems above, with gauge
fixing functions $D^*_A\tau=0$ and $d^*B_3=0,$ respectively.
The covariant gauge functions are given by (\ref{trucm}).

The candidate for the topological action $S_0(A,G)$ is
$\int_{M_{8}} G\wedge G+ \Omega \wedge  \trace (F\wedge F).$  Since
we now have the covariant gauge functions to probe the
moduli space of solutions to (\ref{trucm}) and we have the gauge
fixing functions, we can apply the BRST formalism.
We first express $S_0$ in terms of the norms.  From (\ref{norm}),
\beqa
 \parallel\!
F_A\!\parallel^2
= \int_{M_8} \Omega \wedge \trace (F_A \wedge F_A) 
  +4\parallel\! (F_A)_+\!\parallel^2.
\eeqa  
Also
with $G = G_+ +G_-, G_{\pm}  \in  \Lambda^4_{\pm},$ we have
$\int_{M_{8}} G \wedge G = \parallel\! G_+ \!\parallel^2 -
\parallel\! G_-\!\parallel^2.$  Thus one obtains
 $\parallel\! F_A \!\parallel^2 +
 \parallel\! G\!\parallel ^2 = S_0 + 4 \parallel\!
(F_A)_+ \!\parallel^2 + 2 \parallel\! G_+ \!\parallel^2.$   We know
that $\parallel\! F_A\!\parallel^2$ is minimized when $F_+=0,$
and that $\parallel\! G\!\parallel^2$ is minimized when $G=h.$
So we get a minimum when (\ref{trucm}) is satisfied and it equals
$S_0 + 16\pi^4\alpha^2 \int_{M_{8}}p^2_1=S^1_0.$

In the pure YM case, the natural space was $\LA_P/\LG \times
M_8$ or its subspace $\LM_J \times M_8.$  Rather than 
3-forms on $M_8,$ we need 3-forms on $\LM_J \times M_8$  which
we write as $\widehat{B}_3 = B^0_3 + B^1_2 + B^2_1 + B^3_0$
(eq.(\ref{bghost}), above) with the upper index as the degree in the
$\LM_J$ direction (ghost number) and the lower index in the
$M_8$ direction.  As before $s$ denotes $d_{\LM_{J}}$
so that $(s+d)\widehat{B}_3 = (d_{\LM_{J}} + d_{M_{8}})
\widehat{B}_3 = d_{\LM_{J}} (\widehat{B}_3)
+d_{M_{8}} (\widehat{B}_3)$ is a 4-form with
terms in the $a\choose b$ directions.


\renewcommand{\theequation}{4.\arabic{equation}}\setcounter
{equation}{0}

\section{ BRSTQFT's for other  dimensions than 8 }

 From many points of view the case $D\!=\!8$ is exceptional.
It is of   interest, however, to also  build  BRSTQFT's in other  dimensions, 
by using  the BRST quantization of
d-closed Lagrangians with  gauge functions as in
eq.~(\ref{instp}).
In this section, we first focus on theories with $D\!<\!8$, 
that we directly obtain by various 
dimensional
 reductions in flat space of the J and H theories; we then comment on
the cases
 $D\!=\!12$ and  $D\!=\!10$ .
We will not address the question of observables; their
 determination  is  clear
from the descent equations which can be derived in all
possible cases
from the knowledge of the BRST symmetry.

\subsection{Dimensional reduction of the Yang-Mills 8D
BRSTQFT}

 In $D\!=\!8$, for the J-case, we have seen that there exists a
set of seven
self-duality
equations, on which we have based our BRSTQFT.
These equations were complemented with a Landau gauge
condition to get a
system of 8 independent equations for the 8 components of
$A_\mu$.
These seven equations can be written as
\beq
\Phi_i(F_{\mu\nu}(x^\mu) )=0, \quad 1\leq i\leq 7, \quad
1\leq \mu,\nu\leq 8~.\quad {\rm [  \ (\ref{gaugefix})]}
\eeq

Just as one obtains a BRSTQFT action based on
Bogomolny equations in 3 dimensions \cite{BGM}, we can
define a BRSTQFT in seven dimensions, by standard dimensional reduction
on the eighth
coordinate;
that is,  by putting in the above seven
equations $x^8=0$, $\partial _8=0$   and replacing
$A_8$ by a
scalar field $\varphi(x^j)$ and $F_{i   8} $ by $D_{i}\varphi(x^j)$.
We can then  gauge fix
the longitudinal part of $A_i$, with an  equation of the
following type
\beq
\partial_i A_i=[v, \varphi] ~,
\eeq
which allows for  the case of a massive gauge field $A$.
(Here and in what follows, the constant $v$ defines a direction 
in the Lie algebra for the Yang-Mills symmetry.)
The gauge fixed action will be
\beq\label{red}
\int_{M_7} d^7x\left ( |F_{ij}|^2 + |D_{i }\varphi|^2
+ |\partial_i A_i-[v, \varphi] \ |^2 +{\rm supersymmetric \
terms}\right ) ~.
\eeq

This process can be iterated.
We can go down from  dimension $8$ to $8-n$, by
suppressing  the dependence on $n$ of the coordinates $x^\mu$.
In $D<8$ dimensions we will have a gauge field with $D=8-n$
components and
a set of $n$ scalar fields $\varphi^p$, $p=1,\ldots, n$ which should be
considered as Higgs fields.
Obviously, the dimensional reduction applies as well to the various  ghosts,
and the   fields $\varphi^p$ fall into topological  BRST
multiplets, which, depending on the case, can   possibly be
interpreted as twisted Poincar\'e supermultiplets.  Moreover, 
as we will see when $N=4$,   
there is an interesting option to assign the fields $\varphi^p$
 as elements of other representations, e.g.  spinorial ones, of $SO(D)$.

 One can also consider  
the dimensional reduction in  the  H-case.
One can  break
the symmetry between the  coordinates $y,z,t,w$ and their
complex conjugates by replacing  some  of the fields,
e.g. ${\rm {Im} }~A_{\bar w}$, by   scalar fields.  

In all cases, the final theories rely on
8 independent 
gauge conditions for all fields:  7 for the
topological gauge
ones plus 1
for the ordinary gauge condition, if one starts from the J
case; or 6 for the
3  complex topological gauge conditions plus 2  for the ordinary
 complex  gauge condition,
if one starts from the  H case.

 \subsubsection{The case D=6}

Since the case $D=6$ is of great interest in superstring
theory, let us display
what we  get, starting from the H case,
\beq
\epsilon _{\bar i \bar j \bar k}
 F _{\bar j\bar k}
=D_{\bar i} \varphi ~,
\label{6D}
\eeq
\beq
\partial  _{\bar i }
A _{\bar i}
=[v,\varphi] ~. \label{6Dsub}
\eeq
This set of gauge functions represents 4 complex equations, 
for eight degrees of freedom
represented by the complex fields $A _{\bar i}$ and $\varphi$.

If we start  from the J case, we   have
\beq
\Phi_i(F_{\mu\nu}(x^\mu), D_\mu \varphi^a (x^\mu))=0, \quad
1\leq a \leq 2, \quad
1\leq
\mu,\nu\leq 6 ~,
\eeq
possibly complemented by
\beq
\partial  _{\bar i }
A _{\bar i}
=M_{a,b}\varphi^a\varphi^b+ N_{a,b}v^a\varphi^b~.
\eeq

Notice that
a 2-form gauge field, subjected to the topological
invariance
$sB_2=\Psi_2+\ldots$ can be introduced, still in 6
dimensions,
with the topological self-dual gauge condition
\beq
dB_2+*(dB_2)+\alpha \trace (AdA+{2\over 3}AAA)=0~.
\eeq
This possibility is similar to the introduction of a
3-form in $D=8$.

We can directly build a BRSTQFT in 6 dimensions.
First we consider a pure Yang-Mills case,
taking the topological gauge fixing condition of the type;
\beq
\lambda F_{\mu\nu}~=~\frac{1}{2} T_{\mu\nu\rho\sigma}
F^{\rho\sigma}~.
\label{6dinst}
\eeq
The fourth rank tensor $T_{\mu\nu\rho\sigma}$ is assumed to
be invariant
under some maximal subgroup of $SO(6)$. According to
Corrigan et al \cite{CDFN},
only $SO(4) \times SO(2)$ and $U(3)$ allow
such an invariant tensor. The first choice
corresponds to the case where the 6D manifold is a direct
product of a 4D manifold and 2D Riemann surface; $M_6 =
M_4 \times \Sigma_2$. The second subgroup is
the holonomy group of 6 dimensional K\"ahler manifolds.
In this case we can write down the invariant tensor
as the Hodge dual of a K\"ahler form $\omega$;
\beq
T_{\mu\nu\rho\sigma}~=~(*~\omega)_{\mu\nu\rho\sigma}~. \label{6dtensor}
\eeq
The possible eigenvalues $\lambda$ of
(\ref{6dinst}) with the tensor (\ref{6dtensor}) are $1, -1$
and $-2$.
The eigenspaces of these eigenvalues give the decomposition
of the 15 dimensional representation of $SO(6)$ under its
subgroup
$SU(3) \times U(1)$; ${\bf 15}= {\bf 8} \oplus ({\bf 3}
\oplus \bar {\bf 3})
\oplus {\bf 1}$.\footnote{The usual splitting
of $\Lambda^2 \otimes {\bf C}$ into
 $\Lambda^{1,1} \oplus \Lambda^{2,0} \oplus \Lambda^{0,2}$ with $\Lambda^{1,1}$
decomposed into $\lambda \omega \oplus \omega^\perp$, where $\omega$ is
the K\"ahler form.}  Taking $\lambda =1$
defines the 8 dimensional subspace given by the following
 seven linear conditions on
$F_{\mu\nu}$, where we use complex indices $a,b=1,2, 3$:
\beqa
F_{ab}~=~F_{\bar a \bar b} &=& 0~,  \label{DUY1} \\
\omega^{a \bar b} F_{a \bar b} &=& 0~.
\label{DUY2}
\eeqa
(The last equation (\ref{DUY2}) is, e.g., $F_{1 \bar 1} + F_{2 \bar 2} + F_{3 \bar 3} = 0$.)
The first condition (\ref{DUY1}) means that the connection
is
holomorphic.  These equations are known as 
the Donaldson-Uhlenbeck-Yau (DUY) equation for the moduli space of
stable holomorphic
vector bundles on a K\"ahler manifold.  It  also
appears in
the Calabi-Yau compactification of the heterotic strings.
The DUY equation implies the standard
second order equation of motion for the Yang-Mills 
field\footnote{This is a general property of the system
(\ref{6dinst}).}.
In fact, this follows from the following identity in the
action
density level;
\beqa
-\frac{1}{4} \trace F \wedge * F &+& \omega \wedge \trace (F
\wedge F) \CR
&=& \trace \biggl[
- \frac{3}{2} g^{a\bar a} g^{b\bar b}
F_{a b}F_{\bar a\bar b}
+ (g^{a\bar b} F_{a\bar b})^2  \biggr]~,
\label{6Dnorm}
\eeqa
where we have introduced  the metric $g_{a\bar b}$ for the 
  K\"ahler form $\omega$.
This identity  \cite{BarTian} is crucial in constructing a BRST Yang-
Mills
theory whose classical action is the topological density
$ \omega \wedge \trace (F \wedge F)$.

~From the BRST point of view, one must introduce scalar
 fields to get a correct balance
between the
gauge fixing conditions and the field degrees of freedom 
 and  to recover 
eq.~(\ref{6D}).  
Given a hermitian connection $A$ for the hermitian vector bundle $(E, \rho),$ 
equation (\ref{6D}) says $F^{0,2}_A = (\bar{\partial}_A)^* \widetilde \varphi 
=*^{-1}_1 \partial_A *_1 \widetilde \varphi,$ i.e., $*_1 F^{0,2}_A =
 \partial_A(*_1,\widetilde \varphi) \in \Lambda^{1,3}\otimes \LG.$  
(See section 2.2.1 for the definition of the operation $*_1$.) 
When $M$ is a Calabi-Yau 3 fold, let 
$\varphi = *\widetilde \varphi \in \Lambda^0\otimes \LG$
 and we get $*F^{0,2}_A =
\bar{\partial}_A
\varphi.$ 
 Linearization gives the usual elliptic operator, the holomorphic   of
 $\left(\begin{array}{cc}{\rm curl\ } & {\rm \ grad}\CR
{\rm div} & 0\end{array}\right):$

\beq
\left (\begin{array}{ll}
\bar{\partial}_A & \bar{\partial}^*_A\cr
\bar{\partial}^* _A & 0\end{array}\right)
  :  
\left (\begin{array}{l}
\Lambda^{0,1}\otimes \LG \cr
\Lambda^{0,3}\otimes\LG \end{array}\right) \longrightarrow
\left (\begin{array}{l}
\Lambda^{0,2}\otimes \LG \cr
\Lambda^{0,0}\otimes \LG \end{array}\right)~.
\eeq

Of course what one wants is not (\ref{6D}) but $F^{0,2}_A=0,$ 
equation (\ref{DUY1}), 
the condition that makes $E$ a holomorphic bundle.  
However, as a consequence of the Bianchi identity, 
$\bar{\partial}_A F^{0,2}_A = 0$ and hence (\ref{6D}) implies $\bar{\partial}_A 
\bar{\partial}_A^*\widetilde\varphi = 0,$ which also implies
$\bar{\partial}^*_A \widetilde\varphi=0,$  when $M$ is compact without
boundary.  
Thus (\ref{6D}) implies (\ref{DUY1}); moreover, when $M$ is a Calabi-Yau
3-fold and 
$E$ is stable, $\bar{\partial}^*_A \widetilde\varphi = 0,$
 (equivalent, $\bar{\partial}_A\varphi=0$) only happens when $\varphi$
is  a constant multiple of $I$ in $u(N).$
In that sense, the right hand side of eq.~(\ref{6Dsub}) is 0, giving 
the gauge fixing condition
 $\bar{\partial}* \tau = 0,\ \tau 
\in \Lambda^{0,1}  \otimes \LG$.

Equation (\ref{DUY2}) is the equation $\langle F, \omega\rangle_m=0$ 
(See section 2.3).  
As stated there, the orbit space under complex gauge transformations should be 
the same as symplectic quotient, the orbit space under unitary gauge 
transformations of the 0-momentum set, i.e., the condition 
$\langle F, \omega\rangle_m=0.$
Equation (\ref{6Dnorm}) is a special case of Proposition 3.1 in \cite{BarTian}, 
which we have used previously in Section 2.2.2.

The DUY equation can also be obtained
from the 6 dimensional supersymmetric Yang-Mills theory
on a Calabi-Yau manifold.
The supersymmetry transformation laws of the $(N=1)$ vector
multiplet $(A_M, \Psi)$ in 6 dimensions are
\beqa
\delta A_M &=& i {\overline{\Xi}} \Gamma_M \Psi
- i {\overline{\Psi}} \Gamma_M \Xi~,  \CR
\delta \Psi &=& - \frac{i}{2} \Sigma_{MN} \Xi F^{MN}~,
\eeqa
where $\Gamma_M$ are the gamma matrices and $\Sigma_{MN}
= \frac{1}{4} [ \Gamma_M, \Gamma_N ]$ is the spin
representation.
On the Calabi-Yau manifold the holonomy group is further
reduced to
$SU(3)$, which gives a covariantly constant
(complex) spinor $\zeta$.  In fact this is the very reason
why the
Calabi-Yau manifold is favorable in the compactification of
superstrings to 4 dimensions. We will identify the
supersymmetry
transformation with $\Xi = \zeta$ as a topological BRST
transformation.
With this choice of parameter, SUSY transformations are
decomposed according to the representations of $SU(3)$.
The decomposition of $SO(6)$ vector is
${\bf 6} = {\bf 3} \oplus \bar {\bf 3}$ and the chiral
spinor decomposes
as ${\bf 4} = {\bf 3} \oplus {\bf 1}$.  Thus we obtain the
following
topological BRST transformation law;
\beqa
{s} A_\mu &=& \psi_\mu~,  \quad  {s} A_{\bar\mu} = 0~, \CR
{s} \chi &=& g^{\mu\bar\mu} F_{\mu\bar\mu}~, \quad
{s} \psi_\mu = 0~,  \CR
{s} \psi_{[\bar\mu\bar\nu]} &=& F_{\bar\mu\bar\nu}~, \quad
{s} \rho = 0~.
\eeqa
We should explain how we have \lq\lq twisted\rq\rq  spinors
into
ghosts and anti-ghosts. In terms of the covariantly constant
spinor
$\zeta$ which satisfies $\bar\zeta \Gamma_\mu =0$,
we can make the twist as follows;
\beqa
\chi &=& \bar\zeta \Psi~, \quad
\bar\psi_{\bar\mu} = \bar\zeta \Gamma_{\bar\mu} \Psi~, \CR
 \psi_{[\bar\mu\bar\nu]} &=& \bar\zeta \Gamma_{\bar\mu}
\Gamma_{\bar\nu} \Psi~, \quad \bar\rho = \epsilon^{\bar\mu\bar\nu\bar\sigma}
\bar\zeta \Gamma_{\bar\mu}\Gamma_{\bar\nu} \Gamma_{\bar \sigma }\Psi~,
\eeqa
where $(\bar\psi_{\bar\mu},
\bar\rho)$ are complex conjugates of $(\psi_\mu,
\rho)$.
This is an example of the identification of spinors with forms,
explained in section 2.1.1.
Looking at the BRST transformations of the anti-ghosts,
we recover the DUY equations (\ref{DUY1}, \ref{DUY2}).

\subsubsection{Reduction  to a 4-D BRSTQFT; Seiberg-Witten
equations}

We now turn to the reduction to $D\!=\!4$, which is of special
interest, particularly the theory
obtained by
dimensional reduction of the J theory from $D\!=\!8$ to $D\!=\!4$.
We will get
a BRSTQFT with gauge conditions identical to    the non-Abelian
   Seiberg-Witten equations, which in turn is also related  to the
 $N\!=\!4, D\!=\!4$ supersymmetric theory.

The main observation is that, in the J case   the set of seven
equations (\ref{gaugefix})  can be separated into 3 plus 4
equations. If we group $A_5, A_6,A_7,A_8$ into the 4
component field
$\varphi^\alpha$, $\alpha =1,2,3,4$, the latter can be
interpreted in
4 dimensions as a commuting complex Weyl spinor and  $A_\mu=A_1,
A_2,A_3,A_4$ as a 4 dimensional vector.  The set of the first 3
equations in eq.~(\ref{gaugefix}) can now be interpreted as
the condition that the self-dual part in 4-$D$ of the curvature
of $A_\mu$ is equal to a bilinear in $\varphi^\alpha$; then, the
remaining four
equations can be written as Dirac type equations. 
To be more precise, with the
relevant
definition of the $4\times 4$ matrices $\Gamma_\mu$ and
$\Sigma_{\mu\nu}$,  the dimensional reduction down to $D=4$ of 
eq.~(\ref{gaugefix}) gives
\beqa
F_{\mu\nu}+\epsilon_{\mu\nu\rho\sigma }
F^{\rho\sigma} +^t \varphi  \Sigma _{\mu\nu}\varphi=0~,
\CR
D^{(A)}_\mu \Gamma^\mu \varphi=0~.
\label{gaugefixsw}
\eeqa
The consistency of the dimensional reduction from
eq.~(\ref{gaugefix})
 to eq.~(\ref{gaugefixsw}), and the correctness of the
$SO(4)$
 tensorial properties of all fields, are ensured by the
existence of
 relevant elliptic operators in 8 and 4 dimensions.

The remarkable feature is that the above equations are the
non-abelian version of Seiberg-Witten
equations. In other words, 
we have observed that     the spinors and vectors  of  the non abelian S-W   
theory
get unified in the Yang-Mills field  of the J theory.

The generation of a Higgs potential, to break down the
symmetry, with
a remaining $ U(1)$ is in principle possible, by the
relevant
modifications in the gauge functions, which provide a Higgs
potential,
function of $\varphi$. This is however a subtle issue that
we will
address elsewhere.

The form of the action after dimensional reduction is just
the sum of
 the bosonic part of the Seiberg-Witten action, plus ghost
terms.  Its
 derivation is standard from the knowledge of the gauge
function, as a
 BRST exact term, which enforces the gauge functions.

The link to supersymmetry in 4 dimensions is as follows. The
BRSTQFT
based on $Spin(7)$ is a twisted version of the $D\!=\!8, N\!=\!1$
theory
where the spinor is a complex field counting for $16=8+8$
independent real  components, and
one has a complex scalar field in the supersymmetry
multiplet.  This
theory is itself obtained as the dimensional reduction of
the $D\!=\!10, N\!=\!1$ super Yang-Mills theory, where the spinor has 16
independent
real components. Thus we    predict that the theory   we 
get by dimensional reduction  to 4
dimensions  of   BRSTQFT in 8 dimensions
is related to 
   twisted versions of the $D\!=\!4, N\!=\!4$ super Yang-Mills
theory. For instance, there are 
 6 scalar fields  in the  bosonic sector of the theory as presented 
in the  work of Vafa and Witten \cite{VW},  (see their
eq.(2.1)).
In our derivation, these 6 scalar fields are combinations of
4 of the
components of
the 8-D Yang-Mills field and of  the commuting  ghost and
antighost  $\phi$ and $\bar \phi$ of the J theory.

There are actually three ways of twisting the $N\!=\!4$ SSYM in four
dimensions, 
defined  by how $SO(4) \simeq SU(2) \times
SU(2)$ is embedded in the $R$ symmetry group\footnote{
The $R$ symmetry is the automorphism of  the extended supersymmetry
algebra.} $SU(4)$ \cite{VW}. They are
(i) ({\bf 2}, {\bf 1}) $\oplus$  ({\bf 1}, {\bf 2}) , (ii)  ({\bf 1}, {\bf 2}) 
$\oplus$  ({\bf 1}, {\bf 2}) and (iii)  ({\bf 1}, {\bf 2}) $\oplus$
 ({\bf 1}, {\bf 1}) $\oplus$  ({\bf 1}, {\bf 1}),  where we have indicated
how the defining representation of $SU(4)$ decomposes under
$SU(2) \times SU(2)$.  Taking into account the argument in section 6 of
\cite{BSV},  we can see that the cases (i) and (iii) arise from
the reduction of type H and J cases, respectively.  The remaining
case (ii), which is the twist employed by Vafa-Witten \cite{VW},  is obtained
from the 7 dimensional Joyce manifold with $G_2$ holonomy.
On the other hand, we get the non-abelian Seiberg-Witten theory
with an adjoint hypermultiplet in the case (iii), which
gives the relationship 
between $N=4$ SSYM and non-abelian Seiberg-Witten equation.

We thus conclude that   very interesting  twists  
connect  the fields  of the pure Yang-Mills 8-D BRSTQFT, (obtained 
by gauge fixing the invariant $\O \wedge \trace (F\wedge
F)$),   the fields
which are involved in the four dimensional Seiberg-Witten
equations, and the fields  of the $D\!=\!4, N\!=\!4$ super-Yang-Mills
theory.

We note  that if  one starts from the H case
gauge functions,
the result of compactifying down to 4 dimensions is
 just a complexified version of a two dimensional Yang-Mills
 TQFT, coupled to two scalar fields; it could   also be 
deduced from
 the dimensional reduction of the 3-dimensional BRSTQFT based
on the Bogomolny equations.

 \subsection{Dimensions larger than 8}

\subsubsection{Discussion of the case D=12}

A BRSTQFT  in 12 dimensions  might be a candidate for $F$-theory.
 11-dimensional supergravity,   defined on the boundary
of a 12 dimensional manifold, emphasizes
the relevance of    a 3-form gauge
field $C_3,$ possibly coupled to a non abelian connection
one form $A$.
   The most important  term
$\int_{M_{11}} C_3\wedge dC_3\wedge dC_3 $
of the
 11-dimensional supergravity suggests that
one should
 build a TQFT based on the  gauge-fixing of
 the following invariant \footnote{Here again $dC^3$ means $h+dC^3$ where $h$ is
the harmonic representative of an element in $H^4 (M_{12}).$}
\beq
\int_{M_{12}} \biggl( dC_3\wedge dC_3\wedge dC_3
+dC_3\wedge dC_3 \wedge P_{inv \ 4}(F)
+dC_3 \wedge P_{inv \ 8}(F)
 + P_{inv \ 12}(F) \biggr)~,\eeq
where $ P_{inv \ n}(F)$ are invariant polynomials
of degree $n/2 $ of the curvature
of $A$, i.e,  characteristic classes. 
Special geometries like  hyper or quaternionic
 K\"ahler manifolds give natural four-forms. They, their duals (which are
8-forms, and are therefore good candidates to define gauge functions for the
curvature of a 3-form in 12 dimensions), and their powers might be used as well
here.

 It is natural to try and gauge fix these topological
actions to get a BRSTQFT.  However, we did not  find     gauge fixing
functions
 for a single uncharged 3-form gauge field    in 12 dimensions.  Rather, we did
find one for a single {\it charged} 3-form, and another one 
for a theory with two  {\it uncharged} 3-forms. (See below.)

We could introduce  a 5-form gauge field,
(not relevant for pure 11-dimensional
supergravity),  and similar to the 8-dimensional case,
consider self-duality conditions for the 6-form curvature of
$C_5$,
with a gauge condition of the type
\beq
dC_5+{*{dC_5}}+ \trace (F\wedge F\wedge F) =0~.
\eeq
In the present understanding of superstrings, 5-forms are
not so natural; so
 we will not elaborate further on this case.

When $M_{12}$ is a Calabi-Yau 6-fold, we can
do some things in two different theories. In the first theory, 
we couple a {\it charged} 3-form $B$ to the Yang-Mills
field. ($B$ is valued in the same Lie Algebra as $A$.)
We again use $* :  \Lambda^{0,q} \rightarrow
\Lambda^{0,6-q},$ so that $\Lambda^{0,3} \otimes\LG = \Lambda^{0,3}_+
\otimes\LG + \Lambda^{0,3}_- \otimes \LG.$ 
Again $\bar{\partial}_A
 F^{0,2}-
\bar{\partial}_A\bar{\partial}_A {^*B} = 0$
implies for compact manifolds that 
$\bar{\partial}_A{^*B}=0$.
The covariant gauge
condition is $* F^{0,2} = \bar{\partial}_AB, B \in
\Lambda^{0,3}_+ \otimes \LG;$ equivalently, $F^{0,2} =
\bar{\partial}_A^*B.$ 
So the covariant gauge conditions become  the pair 
$ F^{0,2}=0$ and  
$\bar{\partial}_A B=0$, similar to the Calabi-Yau 3-fold case in section 4.1.1. 
There, $F^{0,2}=0$ and $\bar\partial _A^* \widetilde \varphi =0$, with
$\widetilde \varphi \in  \Lambda ^{0,3}\otimes \LG$. In the present case, 
$B \in  \Lambda_+ ^{0,3}\otimes \LG$.

 The moduli space 
is a vector bundle over the set of
 holomorphic bundles for a fixed $C^\infty$ 
 $(E,\rho)$.
Each such holomorphic structure
gives a unique $A$ with $F_A^{0,2}=0$.
The fiber over
$A$ consists of 
 $[B \in
\Lambda^{0,3}_+\otimes \LG\ ; \ \bar{\partial}_A B=0]$.

The sequence $0 \rightarrow
\Lambda^{0,0} \otimes \LG
\stackrel{\bar{\partial}_A}{\longrightarrow} \Lambda^{0,1}
\otimes\LG \stackrel{\bar{\partial}_A}{\longrightarrow}
\Lambda^{0,2}
\otimes\LG \stackrel{\bar{\partial}_A}{\longrightarrow}
\Lambda^{0,3}_+ \otimes\LG $ is elliptic at the symbol level;
linearization of the covariant gauge condition together with
the usual gauge fixing is given by the elliptic operator:

\beq
\left (\begin{array}{ll}
\bar{\partial}_A & \bar{\partial}^*_A\cr
\bar{\partial}^*_A & 0\end{array}\right)\ :\ 
\left (\begin{array}{l}
\Lambda^{0,1}\otimes\LG \cr
\Lambda^{0,3}_+\otimes\LG \end{array}\right)\longrightarrow
\left (\begin{array}{l}
\Lambda^{0,2}\otimes\LG \cr
\Lambda^{0,0}\otimes\LG \end{array}\right )~.
\eeq

 We take as classical  \lq\lq topological\rq\rq\ action 
$S_0[A,B]=\int_{M_{12}} \Omega_6\wedge  
  \trace (\bar\pa_A  B \wedge  F _A )$ where  $\Omega_6$ is the $(6,0)$
covariant constant form of  $M_{12}$. 
Since the covariant gauge function is 
$F^{0,2}-\bar \partial _A^* B$ and since 
$\langle  F^{0,2} ,\bar \partial _A^* B \rangle =\int_{M_{12}} \Omega_6 \wedge
\trace ( F^{0,2}\wedge \bar \partial _A^* B)$,
we have
$\parallel\! F^{0,2}-\bar \partial _A^* B\!\parallel^2 
= \parallel\! F^{0,2}\!\parallel^2
+ \parallel\!\bar \partial _A^* B\!\parallel^2
- \langle F^{0,2}, \bar \partial _A^* B \rangle
- \langle \bar \partial _A^* B,
F^{0,2} \rangle$, that is, 
$\parallel\! F^{0,2}-\bar \partial _A^* B\!\parallel^2 
= \parallel\! F^{0,2}\!\parallel^2 
+  ~\parallel\!\bar \partial _A^* B\!\parallel^2~
- S_0[A,B] -~^{\bf c}{S_0[A,B]}~
$. (Remember that $\bar \partial _A^*=*\bar \partial _A*$.)
We thus obtain a BRSTQFT whose gauge fixed action will include  the
term  $\parallel\!F^{0,2}\!\parallel^2 +
 \parallel\!\bar \partial _A^* B\!\parallel^2 $. Moreover, the condition  that
$B \in \Lambda^{0,3}_+\otimes \LG\ 
$   can be imposed in a BRST invariant  by using 
the ordinary gauge freedom of $B$ \footnote{The
({0,3})-form 
$B$ is valued in the same Lie   algebra 
  as the Yang-Mills field. It is thus non abelian and 
  its quantization     
  involves the field anti-field formalism of Batalin
   and Vilkoviski. We intend to perform elsewhere  this rather technical task,
which generalizes that sketched at the end of section 3.0.}.

 In the second theory, we introduce {\it two  uncharged}  2-form gauge fields
$B^a_2$ and two (non abelian)  Yang-Mills fields  fields $A^a$, with  $a=1$ and
$2$.  We consider the following   topological 
classical action
\beq\label{classical}
\int_{M_{12}}   \epsilon _{ab}\Omega _6\wedge dB^a_2\wedge dB_2^b~.
\eeq
We define the following   \lq\lq holomorphic\rq\rq\ gauge conditions,
where the complex indices run from 1 to 6
\beq\label{meme}
{^{\bf c}}\partial_{[\bar \mu} B^a_{\bar \nu \bar \rho ]}
+\epsilon ^a_{b}
\epsilon_{\bar \mu \bar \nu \bar \rho \bar \alpha\bar \beta
\bar \gamma }
\partial_{[\bar \alpha} B^b_{\bar \beta \bar
\gamma ]}
= \trace( A^a_{[\bar \mu} \partial_{ \bar \nu}A^a_{\bar
\rho]}
+{2\over 3}
A^a_{[\bar \mu} A^a_{ \bar \nu}A^a_{\bar \rho]} )~.
\eeq
The right hand side of this equation is the Chern-Simons
form of rank 3.
 The   similarity to  8 dimensions is striking, up
to the replacement of
the even Chern class by the odd Chern-Simons class.  
Eq.~(\ref{meme}) implies
\beq
\partial^{\bar \rho}
\partial_{[\bar \mu} B_{\bar \nu \bar \rho ]}^a
=
\epsilon _{b}^a
\epsilon_{\bar \mu \bar \nu \bar \rho \bar \alpha\bar \beta
\bar \gamma }
\trace F^b_{\bar\rho \bar \alpha} F^b_{\bar\beta \bar \gamma
}~.
\label{fin}
\eeq
Its solution  is the stationary point of the following
action
\beq
\label{titi}
\int_{M_{12}} d^{12}x\ 
\epsilon _{ab}
( \partial_{[\bar \mu} B^a_{\bar \nu \bar \rho ]}
{^{\bf c}}{\partial_{[\bar \mu} B^b_{\bar \nu \bar \rho ] }}
+\epsilon_{\bar \mu \bar \nu \bar
\rho \bar \alpha\bar \beta \bar \gamma }
{^{\bf c}}B^a_{\bar \nu \bar \rho }
\trace F^b_{\bar \rho \bar \alpha} F^b_{\beta \bar \gamma }
+{\rm complex \ conjugate})~.
\eeq
Gauge fixing the Lagrangian eq.~(\ref{classical}) by the gauge
condition
eq.~(\ref{fin})  provides a BRST
invariant action. Its ghost independent and gauge independent part  is identical
to the action eq.~(\ref{titi}).

\subsubsection{Other possibilities}

In 10 dimensions
one could build a BRSTQFT based on a four-form  gauge field  $B_4$
and  a pair of two  gauge field 
$B_2^a$, $a=1,2$, which   naturally fit into the type IIB
superstring.  All these forms are uncharged, but they can develop
 non trivial interactions \cite{blb}. The curvatures are
\beq
 G_{5}=dB_4+\epsilon_{ab}B^a_2 G^b_3~,
\eeq
\beq
G_{3}^a=dB^a_2~,  
\eeq
with Bianchi identities, $dG_5=\epsilon_{ab}G^a_3 G^b_3$ and
 $dG^a_3 =0$.
One can construct from  these fields
one closed 11-form
\beq
\Delta_{11}=\epsilon_{ab}G^a_3 G^b_3 G_5~,
\eeq
and two 8-forms
\beq
\Delta_{8}^a=  G_5 G_3  ^a~.
\eeq
The role of the invariant forms  is  obscure, but their existence could
 signal generalizations of Green-Schwarz type anomaly
cancellation mechanism. The   covariant gauge function is
 \beq
dB_4+ *{dB_4}+\epsilon_{ab} B_2^a dB_2^b=0 ~.
\eeq
The mixing of forms of various degrees
   by the gauge functions   generalizes   that 
of  the  3-form with the Yang-Mills field in the
eight dimensional theory of section 3.

\section{Conclusion}

We have described some new Yang-Mills quantum field theories in
dimensions greater than four, using self duality.  In eight dimensions we
found two BRSTQFT's depending on holonomy $Spin(7)$  (the J-case) or
holonomy $SU(4)$ (the H-case).  In the J-case, BRST symmetry is what is left
of  supersymmetry.

The increase in dimension allows us to couple ordinary gauge fields to
forms of higher degree.  We have given several examples.

Dimensional reduction generates new theories.  One of them is a BRSTQFT
whose gauge conditions are the non-abelian Seiberg-Witten equations.

In four dimensions, given the self duality condition, there are other
ways of deriving the  Lagrangian of Witten's topological Yang-Mills
theory besides Witten's twist of $N\!=\!2$ SSYM and besides
BRST \cite{BBRT,CMR,att}.  These methods should work equally well in
deriving our
BRSTQFT Lagrangians for the pure Yang-Mills case.

        Finally, as we have indicated earlier, the geometries of the moduli
spaces we have probed have not been worked out.  Much remains to be
done~\cite{DOT}.  However, from the lessons learned in four dimensions,
it is tempting to hurdle these obstacles and proceed to the
corresponding Seiberg-Witten abelian theory.  Preliminary investigations
 indicate that  one can compute the Seiberg-Witten invariants, when $M_8$
is hyperK\"ahler, i.e., when the holonomy group is $Sp(2).$  This case
is very similar to the Seiberg-Witten invariants for $M_4$ when it is
K\"ahler~\cite{www}.


\vspace{2cm}

{\bf Acknowledgments:}

\bigskip

 H.K.  would like to thank T. Eguchi
and T. Inami for helpful communications.
The work of H.K. is supported in part
by the Grant-in-Aid for Scientific Research
from the Ministry of Education, Science and Culture, Japan.
L. B. would like to thank the Yukawa Institute where part
 of this work has been done, and E. Corrigan and H. Nicolai
for discussions.
IMS would like to thank G.~Tian for bringing him up to date on complex
geometry.  
He also benefited from discussions with S.~Axelrod, S.~Donaldson, R.~Thomas, 
and E.~Weinstein.  His work is supported 
in part by a DOE Grant No. DE-FG02-88ER25066.

 \vspace{2cm}{\bf Note added on July 17, 1997}

 T.A. Ivanova has called our attention to \cite{ivano}, where
 instanton solutions are found.  B.S. Acharya and M.~Loughlin have
 called our attention to their paper \cite {AO} where they discuss
 self duality for Euclidian gravity when $d\leq 8$.  B.S. Acharya,
 M.~Loughlin and B. Spence also discuss self duality in \cite{AOS}. In
 their paper, a note added says that their prood of BRST invariance
 would ``seem to conflict'' with our theory not being topological.
 Indeed the theory is {\it not} topological. They made the corrections
 in a revised version.

We expand on our assertion. Assume $ M$ is a compact oriented simply
connected manifold with $\hat A=1$ and assume $M$ admits a Joyce
metric, i.e, a metric with $Spin(7)$ holonomy. The space of Joyce
metrics modulo diffeomorphisms isotopic to the identity is of
dimension $1+b^4_-(M)$ (see theorem D in \cite{Jo2}). It is conceivable
that this manifold of Joyce metrics is not connected so that one
cannot find a path from one Joyce metric  to another with each point
of the path a Joyce metric.

The BRST argument for invariance requires a path of Joyce metrics,
hence shows formally that the correlation functions are constant on
components of the space of Joyce metrics. But the argument does not
imply constancy of the correlation functions on all Joyce
metrics. This is one reason we chose not to label our J-case QFT a
topological quantum field theory.

On the mathematical side the argument analogous to BRST invariance
also works formally because the correlation functions come from the
second Chern class (See 2.1.3). As we indicated there, to define the
analogue
of Donaldson invariants (the correlation function precisely),
one needs to integrate over the moduli space ${\it M}_J$ of self dual
connections. To do so, a compactification of ${\it M}_J$ is
important. (Work in progress by D.~Joyce and C.~Lewis.)

The H-case (section 2.2.3 in particular) is more
complicated. Physicists allow a degeneration of the complex  structure
to connect one moduli space with another. We do not know how the
``holomorphic Donaldson invariants'' behave under this degeneration.


\end{document}